\documentclass[reprint,
superscriptaddress,
 amsmath,amssymb,
 aps,
]{revtex4-2}

\usepackage{graphicx}% Include figure files
\usepackage{dcolumn}% Align table columns on decimal point
\usepackage{bm}% bold math
\usepackage{hyperref}% add hypertext capabilities
\usepackage{xcolor}
\begin{document}

\preprint{APS/123-QED}

\title{Surface plasmon-phonon-magnon polariton in a topological insulator-antiferromagnetic bilayer structure}

\author{D. Quang To}%
\affiliation{Department of Materials Science and Engineering, University of Delaware, Newark, DE 19716, USA}%

\author{Zhengtianye Wang}
\affiliation{Department of Materials Science and Engineering, University of Delaware, Newark, DE 19716, USA}%

\author{Yongchen Liu}
\affiliation{Department of Materials Science and Engineering, University of Delaware, Newark, DE 19716, USA}%

\author{Weipeng Wu}
\affiliation{Department of Physics and Astronomy, University of Delaware, Newark, DE 19716, USA}%

\author{M. Benjamin Jungfleisch}%
\affiliation{Department of Physics and Astronomy, University of Delaware, Newark, DE 19716, USA}%

\author{John Q. Xiao}%
\affiliation{Department of Physics and Astronomy, University of Delaware, Newark, DE 19716, USA}%

\author{Joshua M.O. Zide}%
\affiliation{Department of Materials Science and Engineering, University of Delaware, Newark, DE 19716, USA}%

\author{Stephanie Law}%
 \email{slaw@udel.edu}
\affiliation{Department of Materials Science and Engineering, University of Delaware, Newark, DE 19716, USA}%

\author{Matthew F. Doty}%
 \email{doty@udel.edu}
\affiliation{Department of Materials Science and Engineering, University of Delaware, Newark, DE 19716, USA}%

\date{\today}% It is always \today, today,

\begin{abstract}
We present a robust technique for computationally studying surface polariton modes in hybrid materials. We use a semi-classical model that allows us to understand the physics behind the interactions between collective excitations of the hybrid system and develop a scattering and transfer matrix method that imposes the proper boundary conditions to solve Maxwell's equations and derive a general equation describing the surface polariton in a heterostructure consisting of N constituent materials. We apply this method to a test structure composed of a topological insulator (TI) and an antiferromagnetic material (AFM) to study the resulting surface Dirac plasmon-phonon-magnon polariton (DPPMP). We find that interactions between the excitations of the two constituents result in the formation of hybridized modes and the emergence of avoided-crossing points in the dispersion relations for the DPPMP. For the specific case of a Bi$_{2}$Se$_{3}$ TI material, the polariton branch with low frequency below 2~THz redshifts upon increasing the thickness of TI thin film, which leads to an upper bound on the thickness of the TI layer that will allow an observable signature of strong coupling and the emergence of hybridized states. We also find that the strength of the coupling between the TI and the AFM, which is parameterized by the amplitude of the avoided-crossing splitting between the two polariton branches at the magnon resonance frequency, depends on the magnitude of the magnetic dipole and the line width of the magnon in the AFM material as well as on the Fermi energy of Dirac plasmon in the TI. Finally, we predict that materials with extremely high quality, i.e. low scattering loss rate, are essential to achieve an experimentally-observable strong coupling between a TI and AFM.
\end{abstract}

\maketitle

\section{Introduction}
Surface Dirac plasmon polaritons (DPP), the electromagnetic collective modes of electrons that are localized evanescent waves in the direction perpendicular to the surface and propagate on the surface of a topological insulator (TI), can be used for a broad range of interdisciplinary applications in sensing, imaging, detection, and photonic data storage in the THz spectral windows \cite{Barnes2003,Pitarke2006,Maier2007,Grigorenko2012,Zhang2012,Garcia2014}. Likewise, magnons, which are the collective excitations of electronic spins in a magnetic material, can be used in realizing high frequency information storage, quantum computing, information transport, and data processing on the micro-scale and nano-scale with extremely low energy consumption owing to the absence of charge transport \cite{Kruglyak2010, Lenk2011,Jungwirth2016,Jungfleisch2018,Lachance2019, Han2019, Pirro2021, Barman2021,Kaffash2021}. If material constituents such as TIs and magnetic materials are combined to form a hybrid material, an incident electromagnetic (EM) wave can excite the internal degrees of freedom of all constituent materials, resulting in the generation of collective excitations (i.e.~polaritons) with emergent properties that provide a possible foundation for novel devices with unique optical and electrical functionalities. For example, the plasmon-magnon interaction can result in a new type of polariton that combines both spin and charge collective excitations into a coherent mode with intriguing and non-trivial properties \cite{Jeong2005, Bludov2019}. The creation and properties of such hybrids has been of interest for some time \cite{Bar1966,Baskaran1973}, but there is still no comprehensive study of such interactions due to the large gap between plasmon and magnon energies in conventional semiconductors or metal systems. Recent advances in the synthesis and fabrication of materials and heterostructures with clean and well controlled interfaces now make it possible to explore the interaction between such excitations. Examples of materials that are now accessible for such studiens include graphene and 3D TIs such as Bi$_{2}$Se$_{3}$, Bi$_2$Te$_3$, Sb$_{2}$Te$_{3}$, all of which host a Dirac plasmon on their surface with energy in the THz spectral window \cite{Ju2011,Pietro2013,Deshk2016,Pietro2020,Wang2020,Chorsi2022}, and antiferromagnetic materials (AFMs) like NiO, MnF$_2$, FeF$_2$ that have magnon energies in the same THz frequency regime \cite{Kotthaus1972,Sanders1981,Rezende2016, Baltz2018, Rezende2019}. 

There have been several prior investigations of the coupling between Dirac plasmons in graphene and magnons in AFMs\cite{Bludov2019,Pikalov2021}. For instance, Bludov and co-authors  used a simple model that neglected all dissipation in the system to study the interaction between a graphene layer and an AFM\cite{Bludov2019}. They found that the dispersion of the surface magnon-plasmon polariton in this system changed drastically upon varying the carrier density of the graphene or the sign of the group velocity. Pikalov and co-workers extended the study of Bludov by taking into account the damping of both the magnons and plasmons\cite{Pikalov2021}. However, to the best of our knowledge, there has not yet been any comprehensive examination of the interaction between a TI and an AFM that considers both electric and magnetic degrees of freedom. Here we present a comprehensive theoretical study of the formation of surface Dirac plasmon-phonon-magnon polaritons in a TI/AFM bilayer structure. Using a semi-classical approach, we investigate the surface polariton modes in a heterostructure by employing the scattering or transfer matrix method to solve Maxwell's equations for an EM wave propagating in the considered system and subject to specific boundary conditions that determine the surface modes. We then derive an analytic equation describing the surface Dirac plasmon-phonon-magnon polariton (DPPMP) in a TI/AFM bilayer associated with the p-polarization of incident light mode, which allows us to explore the properties of the surface DPPMP as a function of the structural parameters of the constituent materials. 

This paper is organized as follows: In Sec. \ref{theo} we describe the methods and models employed in this paper to study the interaction between a TI and AFM. We first review the basic theory starting with Maxwell's equations and standard boundary conditions (Sec. \ref{theo}A). We then give the solutions to Maxwell's equations for the bulk mode within each constituent material (Sec. \ref{theo}B). Using the scattering (or equivalently transfer) matrix technique, we then obtain a general equation for the surface polariton mode in a heterostructure (Sec. \ref{theo}C). In Sec. \ref{result} we apply the method presented in Sec. \ref{theo} to a TI/AFM bilayer structure, beginning with a general consideration of the formation of Dirac plasmon-phonon-magnon polaritons in this system (Sec. \ref{result}A). We then discuss the dependence of these dispersion relations on various combinations of constituent TI and AFM materials and explore the material properties required to obtain an experimentally-observable strong-coupling between the TI and AFM (Sec. \ref{result}B-D). Finally, conclusions and perspectives are provided in Sec. \ref{conc}.

\begin{figure}[h]
\centering
    \includegraphics[width=.4\textwidth]{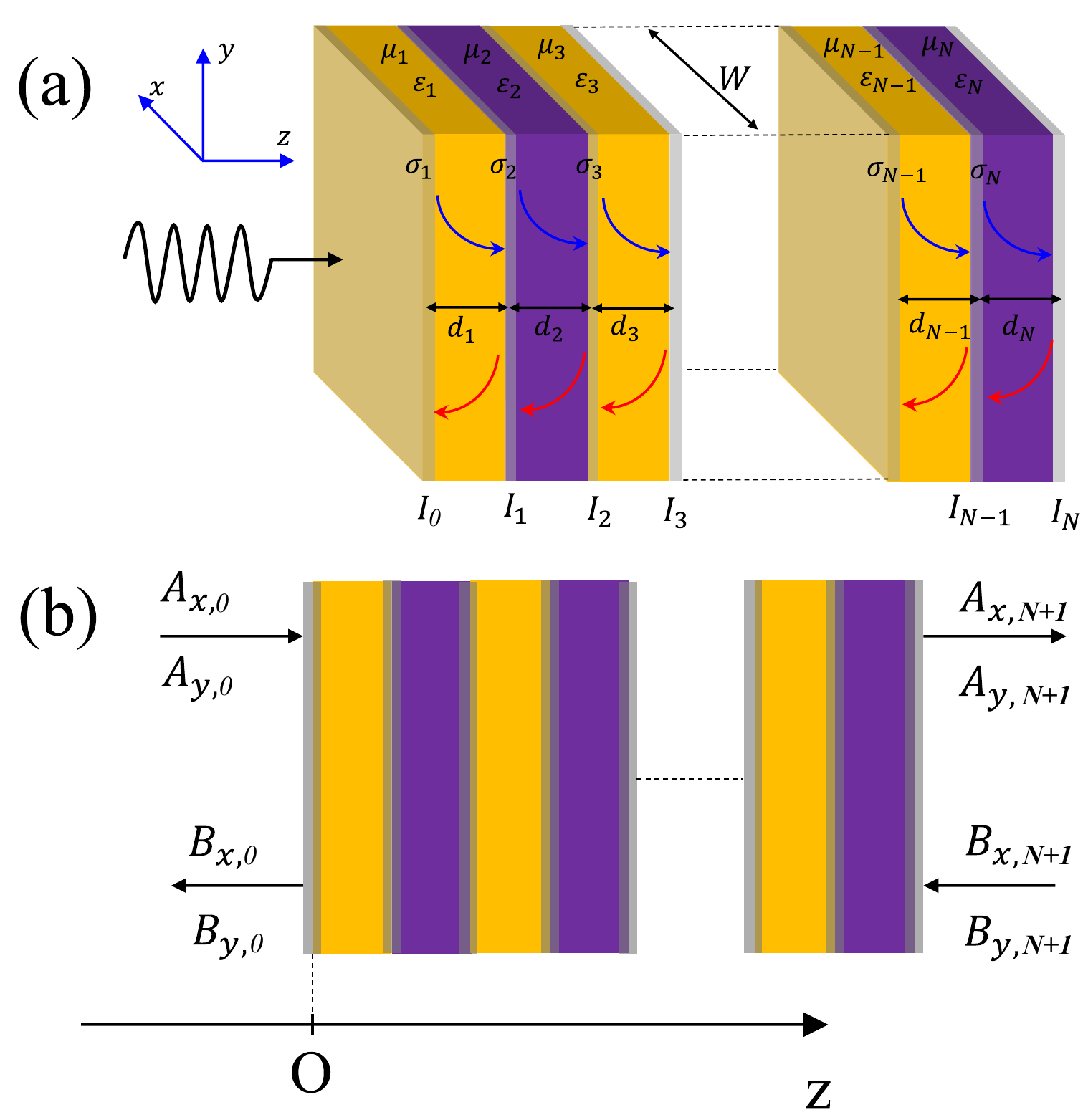}
 \caption{(a) Schematic of a multilayer structure consisting of N constituent layers that have the same width W along the x-direction. The z-axis is chosen as the growth direction of the structure. The thickness, permittivity, permeability in the $m^{th}$ layer, and optical conductivity of the carrier sheet at $m^{th}$ surface/interface are denoted by d$_m$, $\varepsilon_{m}$, $\mu_{m}$ and $\sigma_{m}$, respectively, whereas $I_{m}$ indicates the interface matrix at the $m^{th}$ interface. (b) Schematic of the amplitudes of incoming and outgoing EM waves used in the scattering matrix approach. The EM wave is incident on the left surface in the figure.}
  \label{FIG1}
\end{figure}

\section{Theory}
\label{theo}
Interactions between light and matter can be investigated within three conceptual frameworks \cite{Torma2014,Forndiaz2019}. (1) \textit{Classical description}, in which the collective excitations are considered as harmonic oscillators and their coupling relates to the exchange energy between the two oscillators. In this scheme, the coupling strength between the two oscillators is an input parameter used to fit the dispersion relation $\omega(k)$ to experimental data. For this reason classical models make it difficult to understand the physical origin of the coupling strength or its relationship to the properties of the constituent materials\cite{Sivarajah2019}. (2) \textit{Semi-classical description}, in which Maxwell's equations are used in combination with the optical response functions to describe polaritons. This approach allows one to relate the interaction between two excitations with the structural parameters of the constituent materials through the optical response functions.  (3) \textit{Quantum mechanical representation}, in which the polaritons are hybrid modes, a linear superposition of a matter and photon state. The interaction between the matter and photon states is described through the interacting part of the total Hamiltonian. In all three pictures, the hybridized modes are created when two or more distinct excitations interact with sufficient strength to form new modes that cannot be represented by considering either excitation alone. The signature of hybridized states is an avoided-crossing between the two modes at the point where they would be degenerate in the absence of any interaction. 

Our goal in the present work is to study the coupling between a TI and an AFM that results in the formation of surface Dirac plasmon-phonon-magnon polariton (DPPMP). We then characterize the properties of the surface DPPMP as a function of the structural and material properties of the constituent materials to quantify the coupling between the TI and AFM through the magnetic degree of freedom. These data allow us to understand the physics behind the coupling constant and predict material combinations that might have stronger coupling. To achieve this goal we use the semi-classical approach in which we are going to solve Maxwell's equations to derive the dispersion relationship of the surface DPPMP in a TI/AFM structure. In the following parts of this section we first present a robust technique to obtain the solutions of Maxwell's equation for an EM wave propagating in a heterostructure composed of N constituent materials using state-of-the-art scattering and transfer matrix methods that are computationally effective and capable of dealing with a complex heterostructure \cite{Felbacq1994,Whittaker1999,Li2003,Li2003b,To2022}. Combining these methods with proper boundary conditions describing the surface polariton, an evanescent wave that decays quickly along the propagation direction, we then obtain a general equation determining the surface polariton mode in the heterostructure. We will discuss how to solve this equation numerically in general and then apply it to a specific case with a simple structure involving a TI thin film and an AFM material where an analytical description can be acquired.   

\begin{figure}[h]
\centering
    \includegraphics[width=.5\textwidth]{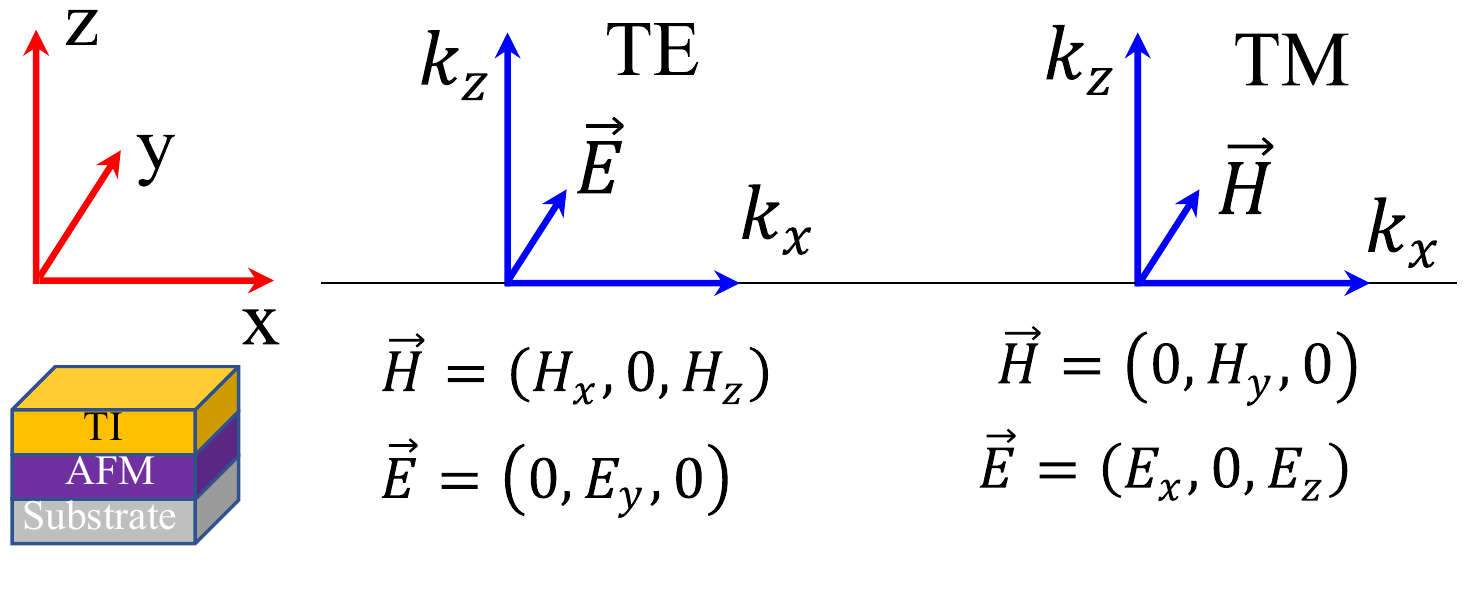}
 \caption{The electric and magnetic components of EM wave corresponding to the TE and TM polarization.}
  \label{FIG2}
\end{figure}

\subsection{Maxwell's equations and boundary condition}
We consider a heterostructure composed of N layers with an EM wave beam incident from the left hand side (see Fig. \ref{FIG1}). We denote the z-axis as the growth direction of the structure. The dimension of the heterostructure along y-direction is infinite while along the x-direction it is finite with a width $W$, as depicted in Fig. \ref{FIG1}. As depicted in Fig. \ref{FIG2}, we set the direction of propagation of the EM wave in the system to be parallel to the x-z plane along the positive direction so that for s-polarization the electric field of the EM wave is polarized along the y-axis. For p-polarization the magnetic field of the EM wave is polarized along the y-direction.

The EM wave propagating within each part of the structure is a solution of Maxwell's equations subject to the standard EM boundary conditions at the interfaces between two materials. In the absence of free volume currents and charges, Maxwell's equations read \cite{Jackson1998,Zangwill2013}:

\begin{align}
    \boldsymbol{\nabla \cdot D} &=0 \label{maxwell1} \\
    \boldsymbol{\nabla} \times \boldsymbol{E}&=-\frac{\partial \boldsymbol{B}}{\partial t} \label{maxwell2}\\
    \boldsymbol{\nabla \cdot B}&=0 \label{maxwell3}\\
    \boldsymbol{\nabla} \times \boldsymbol{H}&=\frac{\partial \boldsymbol{D}}{\partial t} \label{maxwell4}
\end{align}
with
\begin{align}
    \boldsymbol{D} = \varepsilon_{0}\varepsilon \boldsymbol{E} \label{maxwell5}\\
    \boldsymbol{B} = \mu_{0}\mu \boldsymbol{H} \label{maxwell6}
\end{align}
where $\boldsymbol{E}$, $\boldsymbol{D}$, $\boldsymbol{B}$, and $\boldsymbol{H}$ are the electric, displacement, magnetic, and  magnetizing field of the EM wave, respectively; $\varepsilon_{0}$ and $\mu_{0}$ are the permittivity and permeability of free space, respectively; $\varepsilon$ and $\mu$ are, respectively, the relative permittivity and permeability of the media. The boundary conditions at the $m^{th}$ interface are given by:
\begin{align}
\boldsymbol{n}\times \left.\left( \boldsymbol{H}_{m+1} - \boldsymbol{H}_{m}\right)\right\vert_{z=z_{m}} = \boldsymbol{J}_{m} \\
\boldsymbol{n}\times \left.\left( \boldsymbol{E}_{m+1} - \boldsymbol{E}_{m}\right)\right\vert_{z=z_{m}} = 0 \\
\boldsymbol{n} \left.\left( \boldsymbol{D}_{m+1} - \boldsymbol{D}_{m}\right)\right\vert_{z=z_{m}} = \rho_{m} \\
\boldsymbol{n} \left.\left( \boldsymbol{B}_{m+1} - \boldsymbol{B}_{m}\right)\right\vert_{z=z_{m}} = 0
\end{align}
where $\boldsymbol{n}$ is a unit vector perpendicular to the $m^{th}$ interface, $\boldsymbol{J}_{m}$ is the in-plane current, and $\rho_{m}$ is the carrier density of the electron gas at the $m^{th}$ interface.

Taking the curl of equations \ref{maxwell2} and \ref{maxwell4} and inserting \ref{maxwell5} into \ref{maxwell1} and \ref{maxwell6} into \ref{maxwell3}, one obtains:
\begin{align}
    \frac{\mu \varepsilon}{c^{2}}\frac{\partial^{2} \boldsymbol{E} }{\partial t^{2}} + \boldsymbol{\nabla}\left(\boldsymbol{\nabla} \boldsymbol{E} \right)- \boldsymbol{\nabla}^{2}\boldsymbol{E} = 0  \label{1steq} \\
    \boldsymbol{\nabla} \left(\varepsilon_{0}\varepsilon\boldsymbol{E} \right) = 0 \label{2ndeq} \\
    \frac{\mu \varepsilon}{c^{2}}\frac{\partial^{2} \boldsymbol{H} }{\partial t^{2}} + \boldsymbol{\nabla}\left(\boldsymbol{\nabla} \boldsymbol{H} \right)- \boldsymbol{\nabla}^{2}\boldsymbol{H} = 0 \label{3rdeq} \\
    \boldsymbol{\nabla} \left(\mu_{0}\mu\boldsymbol{H} \right) = 0 \label{4theq}
\end{align}
These are the wave equations we solve to obtain the dispersion relationship between the energy (frequency) of the EM wave in the material and the wave vector. In the next section we will give detailed solutions for these equations to derive the bulk polariton mode in each constituent material, i.e.~the TI and AFM. 

\subsection{Bulk polariton modes}
We now consider the bulk polariton modes within each constituent material of the heterostructure displayed in Fig. \ref{FIG1}. Because the propagation direction of the EM wave lies in the x-z plane, within the $m^{th}$ bulk material shown in Fig. \ref{FIG2} the solutions to the wave equations \ref{1steq}, \ref{2ndeq}, \ref{3rdeq},\ref{4theq} can be explicitly written as:
\begin{widetext}
\begin{align}
\boldsymbol{E}_{m} = \begin{pmatrix}
E_{x,m}\\
E_{y,m} \\
E_{z,m}
\end{pmatrix} = e^{i\left(k_{x,m}x - \omega t \right)}\begin{bmatrix} e^{ik_{z,m}z} &0 &e^{-ik_{z,m}z} &0 \\
0 &e^{ik_{z,m}z} &0 &e^{-ik_{z,m}z} \\
-\frac{k_{x,m}}{k_{z,m}}e^{ik_{z,m}z} &0 &\frac{k_{x,m}}{k_{z,m}}e^{-ik_{z,m}z} &0
\end{bmatrix} \begin{pmatrix} A_{x,m} \\
A_{y,m} \\
B_{x,m} \\
B_{y,m}
\end{pmatrix}
\label{Electric}
\end{align}
\end{widetext}
and
\begin{widetext}
\begin{align}
\boldsymbol{H}_{m} = \begin{pmatrix}
H_{x,m}\\
H_{y,m} \\
H_{z,m}
\end{pmatrix} =  \frac{e^{i\left(k_{x,m}x - \omega t \right)}}{\mu_{0}\mu_{m}}\begin{bmatrix} 0 &-\frac{k_{z,m}}{\omega}e^{ik_{z,m}z} &0 &\frac{k_{z,m}}{\omega}e^{-ik_{z,m}z} \\
\frac{1}{\omega}\left(\frac{k_{x,m}^{2}+k_{z,m}^{2}}{k_{z,m}} \right)e^{ik_{z,m}z} &0 &-\frac{1}{\omega}\left(\frac{k_{x,m}^{2}+k_{z,m}^{2}}{k_{z,m}} \right)e^{-ik_{z,m}z} \\
0 &\frac{k_{z,m}}{\omega}e^{ik_{z,m}z} &0 &\frac{k_{z,m}}{\omega}e^{-ik_{z,m}z}
\end{bmatrix} \begin{pmatrix} A_{x,m} \\
A_{y,m} \\
B_{x,m} \\
B_{y,m}
\end{pmatrix}
\label{Magnetic}
\end{align}
\end{widetext}
where $A_{(x,y),m}$ and $B_{(x,y),m}$ are the amplitudes of the x- and y- components of forward- and backward-propagating EM waves, respectively, $\omega$ is the frequency of the EM wave, $k_{x,m}$ and $k_{z,m}$ are the x- and z-components of the wave vector of the EM wave within the $m^{th}$ layer, and x and z are the coordinates along the x- and z- directions. 

Substituting Eq. \ref{Electric} into Eq. \ref{1steq}, and after some algebra, one obtains:
\begin{equation}
    \left[\left(k_{x,m}^{2}+k_{z,m}^{2} \right) \mathcal{I} - \frac{\varepsilon_{m}\mu_{m}\omega^{2}}{c^{2}}  \right] \begin{pmatrix}
    X \\
    Y \\
    Z
    \end{pmatrix} = 0
\end{equation}
which possesses nontrivial solutions when
\begin{equation}
    det\left\vert k_{m}^{2} \mathcal{I} - \frac{\varepsilon_{m}\mu_{m}\omega^{2}}{c^{2}}  \right\vert = 0
    \label{Bulkmode}
\end{equation}
where 
\begin{widetext}
\begin{equation}
   \begin{pmatrix}
    X \\
    Y \\
    Z
    \end{pmatrix}= \begin{bmatrix} e^{ik_{z,m}z} &0 &e^{-ik_{z,m}z} &0 \\
0 &e^{ik_{z,m}z} &0 &e^{-ik_{z,m}z} \\
-\frac{k_{x,m}}{k_{z,m}}e^{ik_{z,m}z} &0 &\frac{k_{x,m}}{k_{z,m}}e^{-ik_{z,m}z} &0
\end{bmatrix} \begin{pmatrix} A_{x,m} \\
A_{y,m} \\
B_{x,m} \\
B_{y,m}
\end{pmatrix}
\end{equation}
\end{widetext}
Here $k_{m}=\sqrt{k_{x,m}^{2}+k_{z,m}^{2}}$ is the total wave vector of the EM wave, $\mathcal{I}$ is the identity matrix, and $\varepsilon_{m}$ and $\mu_{m}$ are, respectively, the dielectric function and magnetic permeability tensors associating with the $m^{th}$ material in the layered structure. In general, one can numerically  solve equation \ref{Bulkmode} using the eigenvalue algorithm to obtain the dispersion of bulk polariton mode $\omega(k_{m})$ of an EM wave propagating within the $m^{th}$ layer of the considered system for an arbitrary magnetic configuration (i.e.~arbitrary $\mu_{m}$) in the magnetic material. 

\begin{figure}[htb]
\centering
    \includegraphics[width=.35\textwidth]{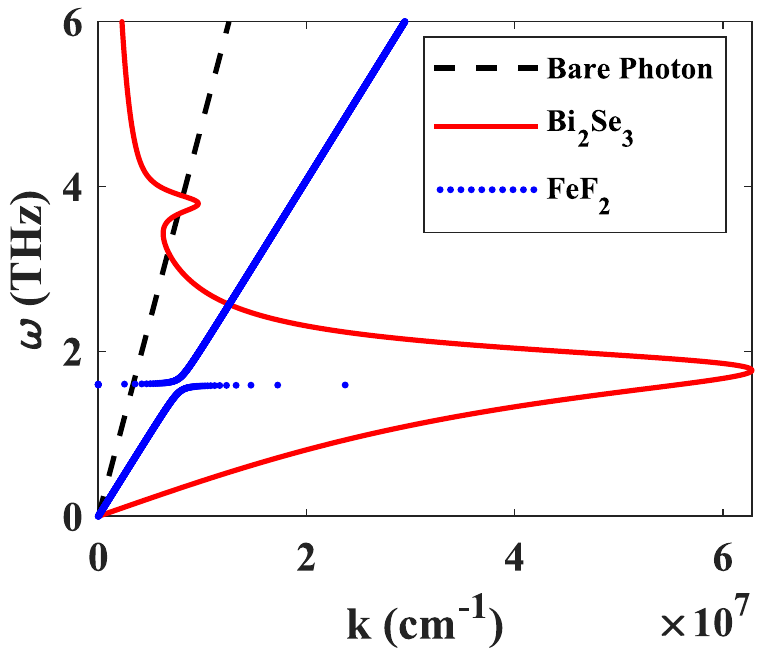}
 \caption{The dispersion of bulk Dirac plasmon-phonon polariton mode in  Bi$_2$Se$_3$ (red curve) and bulk magnon polariton in FeF$_2$ (blue line); the dashed black line represents the bare photon's dispersion relationship $\omega=ck$. The magnetization of the AFM material (FeF$_2$) is along the x-axis.}
  \label{FIG8}
\end{figure}

In figure \ref{FIG8} we plot the dispersion relationship of bulk polariton mode $\omega\left( k\right)$ corresponding to a bare TI (Bi$_2$Se$_3$, red curve) and a bare AFM (FeF$_2$, blue line) associated with TM-polarization and compare them to the bare photon mode (dashed black line). 
For FeF$_2$ we assume that the magnetization is along the x-axis. We see the signature of a magnon collective excitation in the dispersion relationship of bulk magnon polariton mode in FeF$_2$ by the presence of a clear anti-crossing point near the magnon frequency 1.59 THz. The anticrossing is visible due to the low scattering loss rate (small line width) of the magnon in FeF$_2$. We note that we do not directly see a mode associated with the magnon itself (i.e.~constant at 1.59 THz), rather we see the anticrossing in the dispersion relation of bulk magnon polariton mode only when the EM wave is nearly degenerate in energy with the magnon. In the same way, the $\alpha$ and $\beta$ phonons in Bi$_2$Se$_3$ cause the kinks in the dispersion relation of bulk Dirac plasmon-phonon polariton mode of Bi$_2$Se$_3$ around 2 and 4 THz. The wave vector, and hence the momentum, of the bulk magnon polariton mode in FeF$_2$ is always larger than that of light in this regime. Furthermore, as a consequence of weak coupling between the EM wave and the magnon in FeF$_2$ the dispersion of bulk magnon polariton in the FeF$_2$ is almost linear. In contrast, the strong interaction with the  $\alpha$ phonon \cite{To2022} causes the wave vector of the bulk Dirac plasmon-phonon polariton in the Bi$_2$Se$_3$ at low frequency to increase dramatically from 0 to $6 \times 10^{7}~cm^{-1}$, then decrease down to $1 \times 10^{6}~cm^{-1}$ before reaching the bare photon's dispersion, i.e. $k=\frac{\omega}{c}$ at very high frequency (not show).

In a simple picture, one could expect that when a TI and AFM (e.g.~Bi$_2$Se$_3$ and FeF$_2$) are put together to make a hybrid material, the EM wave can interact with both the electric and magnetic excitations in each constituent material via electric and magnetic dipoles. As a result, the dispersion of the EM wave in the hybrid structure will be totally different from those of either of the bare materials. In the next part we will develop a mathematical tool for investigating the surface polariton in the generic heterostructure depicted in Fig. \ref{FIG1}. We then apply it to study the surface DPPMP in a TI/AFM bilayer. 

\subsection{Surface modes}
Let us now turn to the study of surface polaritons, which is the main goal of this work. Starting with the standard boundary conditions for the EM wave at the $m^{th}$ interface of the structure shown in Fig. \ref{FIG1} we have:
\begin{align}
    \left.\boldsymbol{n} \times \left( \boldsymbol{E}_{m+1} -\boldsymbol{E}_{m} \right) \right\vert_{m} = 0 \label{boundary1} \\
    \left.\boldsymbol{n} \times \left( \boldsymbol{H}_{m+1} -\boldsymbol{H}_{m} \right) \right\vert_{m} = \boldsymbol{J}_{m} \label{boundary2}
\end{align}
where
\begin{equation}
    \boldsymbol{n} = \begin{pmatrix}
    0\\
    0\\
    1
    \end{pmatrix}, ~ ~\boldsymbol{J}_{m} = \sigma_{m}\boldsymbol{E}_{m+1}, ~ ~ \sigma_{m} = \begin{pmatrix}
    \sigma_{m}^{xx} &\sigma_{m}^{xy}\\
    \sigma_{m}^{yx} &\sigma_{m}^{yy}
    \end{pmatrix}
\end{equation}
Here $\sigma_{m}$ is the optical conductivity tensor of corresponding two-dimensional carrier gas at $m^{th}$-interface. Substituting Eqs. \ref{Electric} and \ref{Magnetic} into Eqs. \ref{boundary1} and \ref{boundary2}, one obtains
\begin{equation}
    \begin{pmatrix}
    A_{x,m} \\
    A_{y,m} \\
    B_{x,m} \\
    B_{y,m}
    \end{pmatrix} = I_{m}\begin{pmatrix}
    A_{x,m+1} \\
    A_{y,m+1} \\
    B_{x,m+1} \\
    B_{y,m+1}
    \end{pmatrix}
\end{equation}
where $I_{m}$ is an interface matrix that relates the amplitudes of the EM wave in the adjacent $m^{th}$ and $(m+1)^{th}$ layers. If we define:
\begin{equation}
    U = \begin{pmatrix}
    1 &0 &1 &0 \\
    0 &1 &0 &1
    \end{pmatrix}, ~ ~ ~ ~ V = \begin{pmatrix}
    1 &0 &0 \\
    0 &1 &0
    \end{pmatrix}
\end{equation}
then the interface matrix $I_{m}$ will read:
\begin{equation}
    I_{m} = \begin{pmatrix}
    U \\
    L_{m}
    \end{pmatrix}^{-1} \begin{pmatrix}
    U \\
    R_{m}
    \end{pmatrix}
\end{equation}
where 
\begin{equation}
    L_{m}= \frac{V}{\mu_{0}\mu_{m}} \begin{pmatrix}
    0 &-\frac{k_{z,m}}{\omega} &0 &\frac{k_{z,m}}{\omega} \\
    \frac{k_{x,m}^{2}+k_{z,m}^{2}}{\omega k_{z,m}} &0 &-\frac{k_{x,m}^{2}+k_{z,m}^{2}}{\omega k_{z,m}} &0 \\
    0 &\frac{k_{x,m}}{\omega} &0 &\frac{k_{x,m}}{\omega}
    \end{pmatrix}
     \label{Leftmatrix}
\end{equation}
and
\begin{widetext}
\begin{equation}
    R_{m}=  \frac{V}{\mu_{0}\mu_{m+1}} \begin{pmatrix}
    0 &-\frac{k_{z,m+1}}{\omega} &0 &\frac{k_{z,m+1}}{\omega} \\
    \frac{k_{x,m+1}^{2}+k_{z,m+1}^{2}}{\omega k_{z,m+1}} &0 &-\frac{k_{x,m+1}^{2}+k_{z,m+1}^{2}}{\omega k_{z,m+1}} &0 \\
    0 &\frac{k_{x,m+1}}{\omega} &0 &\frac{k_{x,m+1}}{\omega}
    \end{pmatrix} + \begin{pmatrix}
    -\sigma_{m}^{yx} &-\sigma_{m}^{yy} &-\sigma_{m}^{yx} &-\sigma_{m}^{yy} \\
    \sigma_{m}^{xx} &\sigma_{m}^{xy} &\sigma_{m}^{xx} &\sigma_{m}^{xy}
    \end{pmatrix}
     \label{Rightmatrix}
\end{equation}
\end{widetext}
and the propagation matrix between the $m^{th}$ and $(m+1)^{th}$ interfaces is defined by
\begin{widetext}
\begin{equation}
    P_{m,m+1} = \begin{pmatrix}
    e^{-ik_{z,m+1}d_{m+1}} &0 &0 &0 \\
    0 &e^{-ik_{z,m+1}d_{m+1}} &0 &0 \\
    0 &0 &e^{ik_{z,m+1}d_{m+1}} &0 \\
    0 &0 &0 &e^{ik_{z,m+1}d_{m+1}}
    \end{pmatrix}
\end{equation}
\end{widetext}

The amplitudes of EM waves in the outer and inner region can be related by \cite{To2022}
\begin{align}
    \begin{pmatrix}
    A_{0} \\
    B_{0}
    \end{pmatrix} = I_{0}P_{0,1}I_{1} \dots P_{N-1,N}I_{N} \begin{pmatrix}
    A_{N+1} \\
    B_{N+1}
    \end{pmatrix}
\end{align}
The transfer matrix is then defined as:
\begin{equation}
    T= I_{0}P_{0,1}I_{1} \dots P_{N-1,N}I_{N} = \begin{bmatrix} T_{11} &T_{12} \\ T_{21} &T_{22} \end{bmatrix} 
\end{equation}
where $T_{i,j}$ $(i,j = 1,2)$ is itself a $2 \times 2$ matrix and

\begin{align}
    A_{m} = \begin{pmatrix}
    A_{x,m} \\
    A_{y,m}
    \end{pmatrix}, ~ ~ ~ ~ ~ ~ ~ ~ ~ B_{m} = \begin{pmatrix}
    B_{x,m} \\
    B_{y,m}
    \end{pmatrix}
\end{align}

In the same manner, we define a scattering matrix $S_{0,m}$ that connects the amplitudes of the EM wave on the left side of the $0^{th}$ surface (denoted by $I_{0}$ in Fig. \ref{FIG1}) to those on the right side of the $m^{th}$ interface (denoted by $I_{m}$ in Fig. \ref{FIG1}) of the structure:
\begin{equation}
    \begin{pmatrix}
    A_{m}\\
    B_{0}
    \end{pmatrix} = S_{0,m} \begin{pmatrix}
    A_{0}\\
    B_{m}
    \end{pmatrix}
\end{equation}
and a transfer matrix $M_{m,l}$ that connects the amplitudes of the EM wave on the right side of the $m^{th}$ interface to those on the right side of the $l^{th}$ interface ($l>m$) of the structure:
\begin{equation}
    \begin{pmatrix}
    A_{m}\\
    B_{m}
    \end{pmatrix} = M_{m,l} \begin{pmatrix}
    A_{l}\\
    B_{l}
    \end{pmatrix}
\end{equation}
Then the scattering matrix $S_{0,l}=S_{0,m} \otimes M_{m,l}$ that relates the outgoing and incoming states on the left side of the $0^{th}$ surface to those on the right side of $l^{th}$ interface,
\begin{equation}
    \begin{pmatrix}
    A_{l}\\
    B_{0}
    \end{pmatrix} = S_{0,l} \begin{pmatrix}
    A_{0}\\
    B_{l}
    \end{pmatrix}
\end{equation}
will be obtained via a recursive method:
\begin{widetext}
\begin{align}
    S_{0,l}^{11} &= \left[\mathcal{I} - \left(M_{m,l}^{11} \right)^{-1}S_{0,m}^{12}M_{m,l}^{21} \right]^{-1}\left(M_{m,l}^{11} \right)^{-1}S_{0,m}^{11}  \\
    S_{0,l}^{12} &= \left[\mathcal{I} - \left(M_{m,l}^{11} \right)^{-1}S_{0,m}^{12}M_{m,l}^{21} \right]^{-1}\left(M_{m,l}^{11} \right)^{-1}\left(S_{0,m}^{12}M_{m,l}^{22}-M_{m,l}^{12} \right)  \\
    S_{0,l}^{21} &=S_{0,m}^{22}M_{m,l}^{21} S_{0,l}^{11} + S_{0,m}^{21}  \\
    S_{0,l}^{22}&=S_{0,m}^{22}M_{m,l}^{21} S_{0,l}^{12} + S_{0,m}^{22}M_{m,l}^{22} 
\end{align}
\end{widetext}
where $\mathcal{O}_{m,l}^{11}, ~ \mathcal{O}_{m,l}^{12}, ~ \mathcal{O}_{m,l}^{21}, ~ \mathcal{O}_{m,l}^{22}$ are $2\times2$ block elements of the matrix $\mathcal{O}_{m,l}~(\mathcal{O} \equiv S, M)$, $\mathcal{I}$ is the identity matrix. Overall, one can construct the total scattering matrix $S=S_{0,N}=S_{0,1} \otimes M_{1,2} \otimes ... \otimes M_{N-1,N}$ that links the amplitudes of EM waves in the outer and inner region:
\begin{equation}
    \begin{pmatrix}
    A_{N+1}\\
    B_{0}
    \end{pmatrix} = S \begin{pmatrix}
    A_{0}\\
    B_{N+1}
    \end{pmatrix}
    \label{scatt}
\end{equation}
leading to the well-known relationship between the scattering matrix $S$ and the transfer matrix $T$:
\begin{equation}
    S = \begin{bmatrix}
    T_{11}^{-1} & -T_{11}^{-1}T_{12}\\
    T_{21}T_{11}^{-1} & T_{22}-T_{21}T_{11}^{-1}T_{12}
    \end{bmatrix} = \begin{bmatrix}
    t & r' \\
    r & t'
    \end{bmatrix}
\end{equation}
where $t,t'$ and $r,r'$ are each $2 \times 2$ matrices indicating the transmission and reflection coefficients of the total electromagnetic wave which is in general with both TE- and TM-polarization. Here $t$ and $r$ are, respectively, the transmission and reflection associated with an incident wave propagating along the +z direction, whereas $t'$ and $r'$ correspond to the incident wave propagating along the -z direction. This formulation allows for an explicit picture of the reflection and transmission coefficients for the TE and TM polarized EM waves propagating in the structure.

In order to apply these techniques to investigate the surface polariton, we note that a key feature of the surface polariton mode is that it is an evanescent wave that carries energy propagating laterally, i.e , in the x-direction in our coordinate system, while decaying exponentially in the z direction. Without loss of generality, we consider a Cartesian coordinate system as depicted in Fig. \ref{FIG1}a with the origin O at the left surface of the structure (see Fig \ref{FIG1}b). This leads to the condition $A_{0}=0$ for the surface polariton excited on the surface of the structure corresponding to the interface matrix $I_{0}$ indicated in Fig. \ref{FIG1}. Further, at the right hand side of the terminated surface of the structure, i.e the surface associated with interface matrix $I_{N}$ in Fig. \ref{FIG1}, the EM wave is strictly outgoing wave (i.e.~nothing is incident from the right). This yields $B_{N+1}=0$. With these boundary conditions, the amplitudes of the EM wave determining the surface polariton mode of the heterostructure shown in Fig. \ref{FIG1}, in the transfer matrix formalism, are given by:
\begin{align}
    \begin{pmatrix}
    0 \\
    B_{0}
    \end{pmatrix} = \begin{bmatrix} T_{11} &T_{12} \\ T_{21} &T_{22} \end{bmatrix} \begin{pmatrix}
    A_{N+1} \\
    0
    \end{pmatrix}
\end{align}
which possesses non-trivial solutions only if
\begin{equation}
    det \left[ T_{11} \right] = 0
    \label{Surfacemode1}
\end{equation}

In term of the scattering matrix, we re-write Eq. \ref{scatt} as:
\begin{equation}
    S^{-1}\begin{pmatrix}
    A_{N+1}\\
    B_{0}
    \end{pmatrix} =  \begin{pmatrix}
    A_{0}\\
    B_{N+1}
    \end{pmatrix}
\end{equation}
Applying the condition $A_{0}=B_{N+1}=0$, we obtain the condition for non-trivial solutions:
\begin{equation}
    \frac{1}{det\vert S \vert} =0
    \label{Surfacemode2}
\end{equation}

Equations \ref{Surfacemode1} and \ref{Surfacemode2} are general equations that allow us to determine the surface polariton mode dispersion $\omega \left(k_{x} \right)$ for a general heterostructure with N layers. To solve those equation \ref{Surfacemode1} and \ref{Surfacemode2} numerically, we vary the frequency $\omega$ and the in-plane wave vector $k_{x}$. The surface polariton mode then corresponds to the local maximum of the function $F\left(\omega,k_{x} \right) = \frac{1}{det \vert T_{11} \vert}$ or $F'\left(\omega,k_{x} \right) = det\vert S \vert$. For this reason, a plot of $F$ or $F'$ as a function of $\omega$ and $k_{x}$ accurately represents the dispersion relation. Notably, Eq. \ref{Surfacemode1} allows for a simple analytical derivation of the surface polariton mode in a structure with few layers. Meanwhile, Eq. \ref{Surfacemode2} gives better numeric convergence when there are a large number of layer. This is due to the advantages of the scattering matrix method, which relates the incident and outgoing fields and avoids their mixture \cite{Felbacq1994,Whittaker1999,Li2003,Li2003b}. 

\section{Results and discussions}
\label{result}
We now apply the methods presented in the previous section to investigate the interactions between a TI and an AFM. The input parameters for our calculations are the thickness of the corresponding material constituents, the frequency dependent dielectric functions, the permeability tensors, and the optical conductivities of the two dimensional carriers on the surface and at the interface between two materials. 

\begin{figure}[htb]
\centering
    \includegraphics[width=.5\textwidth]{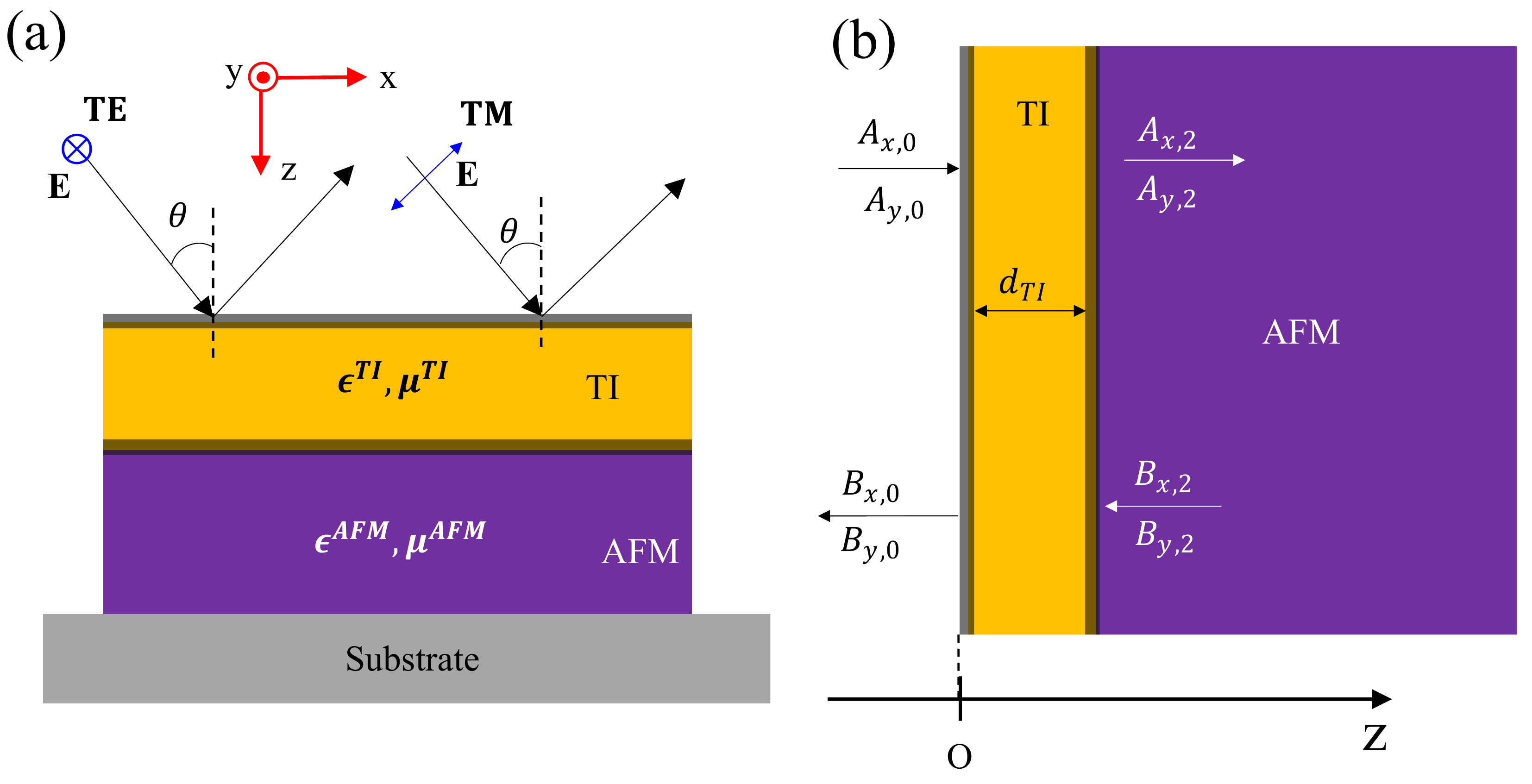}
 \caption{(a) The structure under consideration in the remainder of this work. A TI is deposited on top of an AFM with arbitrary magnetization direction. An EM wave with, in general, both TE- and TM- polarization is incident on the top surface of the TI material to excite the electric degree of freedom in the TI thin film, which can then couple with the magnetic degree of freedom in the AFM layer. (b) A finite TI film with thickness $d_{TI}$ in contact with a half-infinite AFM. $A_{i,j}$ and $B_{i,j}$, with $i \equiv x, y$ and $j \equiv 0,2$, are the amplitudes of the forward- and backward-propagating EM waves in the air (indicated by $j=0$) and within the AFM material (indicated by $j=2$).}
  \label{FIG3}
\end{figure}

In the absence of an external magnetic field, the optical conductivity tensor of the two-dimensional carrier gas on the surface and at the interface between two materials takes a diagonal form:

\begin{equation}
    \boldsymbol{\sigma} = \begin{pmatrix}
    \sigma &0 \\
    0 &\sigma
    \end{pmatrix}
\end{equation}
i.e, $\sigma^{xx} = \sigma^{yy} \equiv \sigma$ and $\sigma^{xy}= \sigma^{yx}=0$. The magnetic permeability tensor of a magnetic layer can be written as \cite{Mills1974}:
\begin{equation}
    \mu = \begin{bmatrix}
    \mu_{xx} &0 &0 \\
    0 & \mu_{yy} &0 \\
    0 &0 &\mu_{zz} 
    \end{bmatrix}
    \label{mu}
\end{equation}
where $\mu_{\xi\xi} = 1$ if the magnetization is along the $\xi$-direction and $\mu_{\xi\xi} = 1+ 4\pi \frac{2\gamma^{2}H_{a}M}{\Omega_{0}^{2}-\left(\omega +i/\tau_{mag} \right)^{2}}$ otherwise. $\xi \equiv x,y,$ or $z$. Here $H_{a}$ is the anisotropy field, $H_{e}$ is the exchange field, $M$ is the sublattice magnetization, $\Omega_{0}=\gamma\sqrt{\left(2H_{e}+H_{a} \right)H_{a}}$ is the resonance frequency, $\gamma=g\frac{e}{2mc}$ is the gyromagnetic ratio in cgs units, $g$ is the Lande g-factor, and $\tau_{mag}$ is the damping constant for the AFM.  

The dielectric function of corresponding layers, which are isotropic materials considered in this work, in the structure is given by the Drude-Lorentz model \cite{Dordevic2013}:
\begin{equation}
    \varepsilon\left( \omega \right) = \varepsilon_{\infty} - \frac{\omega_{p}^{2}}{\omega^{2}+i\Gamma \omega} +  \sum_{n=1}^{N}\frac{\omega_{p,n}^{2}}{\omega_{0,n}-\omega^{2}-i\Gamma_{n} \omega}
    \label{dielectric}
\end{equation}
where $\varepsilon_{\infty}$ is the dielectric constant at high frequency $\left( \omega \rightarrow \infty \right)$, the second term on the right hand side of Eq. \ref{dielectric} indicates the Drude bulk contributions, and the third term is a sum of all contributions from the other Lorentz oscillators presenting in the system. 

\subsection{Surface polariton mode in a TI/AFM structure: General considerations}
We start with a general consideration of the formation of surface Dirac plasmon-phonon-magnon polaritions (DPPMP) in the TI/AFM structure shown in Fig. \ref{FIG3}, where an AFM on a substrate (MgO) is capped with a TI thin film. An EM wave with both TM- and TE- polarization is incident on the top of the TI thin film to excite the collective excitations, specifically (a) the Dirac plasmons on the surface of the TI and at the interface between the TI and AFM and (b) the magnon in the AFM. These electric and magnetic excitations of the system can interact with each other to create new hybrid modes (i.e.~DPPMPs) that manifest as a change in the dispersion relation $\omega(k)$. We analyze the emergence of these DPPMPs below.  

For the sake of simplicity, we assume that the AFM is sufficiently thick to be considered as half-infinite. We then indicate the amplitudes of incoming and outgoing EM wave as denoted in Fig. \ref{FIG3}(b). The surface DPPMP dispersion relationship for this structure can be acquired under the conditions $A_{x,0}=A_{y,0}=B_{x,2}=B_{y,2}=0$, which yield a transfer matrix determined by $T=I_{0}PI_{1}$ where $I_{0}$ ($I_{1}$) is the interface matrix between the air and the TI (TI and AFM) and $P$ is the propagation matrix within the TI thin film. To construct the transfer matrix $T$ for the solutions of surface DPPMP in the system, we first establish the interface matrices $I_{0}$ and $I_{1}$ by inserting Eq. \ref{mu} into Eqs. \ref{Leftmatrix} and \ref{Rightmatrix}. After some algebra, one obtains:
\begin{equation}
    I_{m} = \begin{pmatrix}
    1+\mathcal{A}+\mathcal{B} &0 &1-\mathcal{A}+\mathcal{B} &0 \\
    0 &1+\mathcal{C}+\mathcal{D} &0 &1-\mathcal{C}+\mathcal{D} \\
    1-\mathcal{A}-\mathcal{B} &0 &1+\mathcal{A}-\mathcal{B} &0 \\
    0 &1-\mathcal{C}-\mathcal{D} &0 &1+\mathcal{C}-\mathcal{D}
    \end{pmatrix}
\end{equation}
where
\begin{align}
    \mathcal{A} = \frac{\mu_{yy}^{m}k_{z,m}\left(k_{x,m+1}^{2}+k_{z,m+1}^{2} \right)}{\mu_{yy}^{m+1}k_{z,m+1}\left(k_{x,m}^{2}+k_{z,m}^{2} \right)} \\
    \mathcal{B} = \frac{\mu_{0}\mu_{yy}^{m}k_{z,m}\omega \sigma_{m}}{k_{x,m}^{2}+k_{z,m}^{2}} \\
    \mathcal{C} = \frac{\mu_{xx}^{m}k_{z,m+1}}{\mu_{xx}^{m+1}k_{z,m}} \\
    \mathcal{D} = \frac{\mu_{0}\mu_{xx}^{m}\omega\sigma_{m}}{k_{z,m}}
\end{align}
and $m=0,1$ with
\begin{align}
    \mu_{\xi\xi}^{0}=1 \\
    \mu_{\xi\xi}^{1}=\mu_{\xi\xi}^{TI} \\
    \mu_{\xi\xi}^{2} =\mu_{\xi\xi}^{AFM}
\end{align}
the permeability of the air, TI, and AFM, respectively. Here $\xi \equiv x,y,$ or $z$, the coordinate axes; $\sigma_{0}$ and $\sigma_{1}$ are, respectively, the optical conductivity of the Dirac plasmon on the surface of the TI layer and at the interface between the TI and the AFM. Using the explicit forms of the interface matrices $I_{0}$ and $I_{1}$, together with the propagation matrix
\begin{equation}
    P = \begin{pmatrix}
    e^{-ik_{z,1}d_{TI}} &0 &0 &0 \\
    0 &e^{-ik_{z,1}d_{TI}} &0 &0 \\
    0 &0 &e^{ik_{z,1}d_{TI}} &0 \\
    0 &0 &0 &e^{ik_{z,1}d_{TI}}
    \end{pmatrix}
\end{equation}
one can obtain the transfer matrix for the structure in Fig. \ref{FIG3}:
\begin{equation}
    T=I_{0}PI_{1}=\begin{pmatrix}
    T_{11} &T_{12} \\
    T_{21} &T_{22}
    \end{pmatrix}
\end{equation}

The surface DPPMP modes in the TI/AFM bilayer satisfy
\begin{equation}
    det \left[T_{11} \right] =0
    \label{determ}
\end{equation}
Because we do not consider an external magnetic field, the TE and TM modes are uncoupled, leading to a diagonal form for $T_{11}$:
\begin{equation}
    T_{11} = \begin{pmatrix}
    T_{11}^{11} &0 \\
    0 &T_{11}^{22}
    \end{pmatrix}
\end{equation}

Solutions of Eq. \ref{determ} would correspond to $T_{11}^{11}=0$ and $T_{11}^{22}=0$, associated respectively with the $TM$- and $TE$-polarization of the light incident on the structure. The TE-polarization light cannot excite the Dirac plasmon on the surface of the TI material \cite{To2022}, so we consider only the $TM$-polarization of the EM wave determined by $T_{11}^{11}=0$, which gives:
\begin{widetext}
\begin{align}
\begin{split}
    \left[1+\frac{k_{z,0}\left(k_{x}^{2}+k_{z,1}^{2} \right)}{\mu_{yy}^{TI}k_{z,1}\left(k_{x}^{2}+k_{z,0}^{2} \right)} +\frac{\mu_{0}\omega k_{z,0}\sigma_{0}} {k_{x}^{2}+k_{z,0}^{2}}\right] \left[1+\frac{\mu_{yy}^{TI}k_{z,1}\left(k_{x}^{2}+k_{z,2}^{2} \right)}{\mu_{yy}^{AFM}k_{z,2}\left(k_{x}^{2}+k_{z,1}^{2} \right)}+\frac{\mu_{0}\mu_{yy}^{TI}\omega k_{z,1} \sigma_{1}}{k_{x}^{2}+k_{z,1}^{2}} \right]e^{-ik_{z,1}d_{TI}} \\
    +\left[1-\frac{k_{z,0}\left(k_{x}^{2}+k_{z,1}^{2} \right)}{\mu_{yy}^{TI}k_{z,1}\left(k_{x}^{2}+k_{z,0}^{2} \right)} +\frac{\mu_{0}\omega k_{z,0}\sigma_{0}} {k_{x}^{2}+k_{z,0}^{2}}\right] \left[1-\frac{\mu_{yy}^{TI}k_{z,1}\left(k_{x}^{2}+k_{z,2}^{2} \right)}{\mu_{yy}^{AFM}k_{z,2}\left(k_{x}^{2}+k_{z,1}^{2} \right)}-\frac{\mu_{0}\mu_{yy}^{TI}\omega k_{z,1} \sigma_{1}}{k_{x}^{2}+k_{z,1}^{2}} \right]e^{ik_{z,1}d_{TI}} =0
\end{split}
\label{TM_surface}
\end{align}
\end{widetext}

Eq. \ref{TM_surface} is general because it applies to various kind of TI/AFM bilayer combination. This equation can be solved numerically to obtain the dispersion of the surface polariton in a TI/AFM structure once the optical response function and thickness of the constituent materials are known. We note that for the p-polarization, one has the magnetic field of the EM wave along the y-direction. This suggests that magnetization of the AFM along the y-direction would not exert any impact on the spectrum of surface DPPMPs because the $yy$- component of the permeability tensor will be 1 in this case ($\mu_{yy}^{AFM}=1$). We therefore consider the case in which the magnetization of the AFM is along $x$-direction, i.e. perpendicular to the magnetic field of the EM wave, which yields an AFM permeability of the form:
\begin{equation}
    \mu_{yy}^{AFM} =\mu^{AFM} =1+4\pi\frac{2\gamma^{2}H_{a}M}{\omega_{0}^{2}-\left(\omega+i/\tau_{mag} \right)^{2}}
    \label{a2}
\end{equation}
Solving Eq. \ref{Bulkmode} for the considered configuration, we obtain the bulk modes within each region given by:
\begin{equation}
    k_{z,0}= \sqrt{\frac{\omega^{2}}{c^{2}} -k_{x}^{2}}
    \label{a3}
\end{equation}

\begin{equation}
    k_{z,1}=\sqrt{\frac{\varepsilon^{TI}\mu^{TI}\omega^{2}}{c^{2}} - k_{x}^{2}}
    \label{a4}
\end{equation}

\begin{equation}
k_{z,2}=\sqrt{ \frac{\varepsilon^{AFM} \mu^{AFM}\omega^{2}}{c^{2}} - k_{x}^{2}}
\label{a5}
\end{equation}
Here $k_{x} = q \approx \frac{W}{\pi}$ where $W$ is the width of the TI and AFM ribbon along the x-direction. There is no magnetic order in the TI materials in this consideration, so  
\begin{equation}
   \mu^{TI}=\mu_{yy}^{TI} = 1 
   \label{a1}
\end{equation}
Substituting the relations \ref{a1}, \ref{a2}, \ref{a3}, \ref{a4} and \ref{a5} into Eq. \ref{TM_surface}, we finally obtain:
\begin{widetext}
\begin{align}
 \begin{split}
     F^{-1}= \left(1 + \frac{\varepsilon^{TI}k_{z,0}}{k_{z,1}} + \frac{\sigma_{0}k_{z,0}}{\varepsilon_{0}\omega} \right)\left(1 + \frac{\varepsilon^{AFM}k_{z,1}}{\varepsilon^{TI}k_{z,2}} +\frac{\sigma_{1}k_{z,1}}{\varepsilon_{0}\varepsilon^{TI}\omega} \right)e^{-ik_{z,1}d_{TI}} \\
     + \left(1 - \frac{\varepsilon^{TI}k_{z,0}}{k_{z,1}} + \frac{\sigma_{0}k_{z,0}}{\varepsilon_{0}\omega} \right)\left(1 - \frac{\varepsilon^{AFM}k_{z,1}}{\varepsilon^{TI}k_{z,2}} -\frac{\sigma_{1}k_{z,1}}{\varepsilon_{0}\varepsilon^{TI}\omega} \right)e^{ik_{z,1}d_{TI}}  = 0
\end{split}   
\label{TM_surface_p}
\end{align}
\end{widetext}

In the limit $d_{TI} \rightarrow 0$, Eq. \ref{TM_surface_p} reduces to
\begin{equation}
    1 + \frac{\varepsilon^{AFM}k_{z,0}}{k_{z,2}} + \frac{\left(\sigma_{0}+\sigma_{1} \right)k_{z,0}}{\varepsilon_{0}\omega} =0
\end{equation}
This is an equation that describes the surface Dirac plasmon-magnon polariton in a graphene-like/AFM system \cite{Pikalov2021}. The third term $\frac{\left(\sigma_{0}+\sigma_{1} \right)k_{z,0}}{\varepsilon_{0}\omega}$ on the left hand side includes the contribution from the two surfaces of the TI, which are degenerate in the $d_{TI} \rightarrow 0$ limit. 

\subsection{Parameters and relationships for specific TI/AFM structures}
To investigate the surface DPPMP in specific TI/AFM structures, we consider three TI candidates, Bi$_{2}$Se$_3$, Bi$_{2}$Te$_3$ and Sb$_{2}$Te$_3$, whose bulk dielectric function in the far-IR range of interest is given by \cite{To2022}:
\begin{equation}
    \varepsilon^{TI} = \varepsilon_{\infty} + \frac{S_{\alpha}^{2}}{\omega_{\alpha}^{2}-\omega^{2}-i\omega\Gamma_{\alpha}} + \frac{S_{\beta}^{2}}{\omega_{\beta}^{2}-\omega^{2}-i\omega\Gamma_{\beta}}
\end{equation}
where $\omega_{x}$, $\Gamma_{x}$, and $S_{x}$ are the frequency, the scattering rate, and the strength of the Lorentz oscillator associated with the $\alpha$ ($x=\alpha$) and the $\beta$ ($x=\beta$) phonons of the TI thin film. Numerical values for all TI parameters are taken from reference \cite{Deshk2016} and listed in Table \ref{tableTI}. 
\begin{widetext}
\center
\begin{table}[h]
\caption{The TI parameters used in this work, taken from \cite{Deshk2016}.}
\begin{tabular}{cccccccc}
\hline
\hline
Materials &$\varepsilon_{\infty}$ &S$_{\alpha}$(cm$^{-1}$) &$\omega_{\alpha}$(cm$^{-1}$) &$\Gamma_{\alpha}$(cm$^{-1}$)  &S$_{\beta}$ (cm$^{-1}$) &$\omega_{\beta}$(cm$^{-1}$) &$\Gamma_{\beta}$ (cm$^{-1}$) \\
\hline
Bi$_{2}$Se$_{3}$  &1 &675.9 &63.03 &17.5 &100 &126.94 &10 \\
Bi$_{2}$Te$_3$  &85 &716 &50 &10 &116 &95 &15  \\
Sb$_{2}$Te$_3$  &51 &1498.0 &67.3 &10 &NA &NA &NA\\
\hline
\hline
\end{tabular}
\label{tableTI}
\end{table}\vspace{-15pt}
\center
\begin{table}[h]
\caption{The parameters for AFM materials used in this paper, taken from \cite{Rezende2016,Rezende2019,Pikalov2021}.}
\begin{tabular}{ccccccccc}
\hline
\hline
Materials &$H_{a}$ (Oe) &$H_{e}$ (Oe) &M (G) &$\omega_{0}$ (THz) &$\tau_{mag}$ (ns) &Lande factor &$T_{Neel}$ (K)  \\
\hline
NiO & $6.4\times10^{3}$ &$9.7\times10^{6}$ &400 &1.01 &0.175 &2.05 &523\\
MnF$_2$ &$8\times10^{3}$ &$5.33\times10^{5}$ &592 &0.26 &7.58 &2.0  &67\\
FeF$_2$ &$2\times10^{5}$ &$5.4\times10^{5}$ &560 &1.59 &0.11 &2.25 &78\\
\hline
\hline
\end{tabular}
\label{tableafm}
\end{table}
\end{widetext}

The surface states of these TIs host two dimensional spin-polarized (Dirac) plasmons that behave as a conducting electron sheet whose optical conductivity is given by
\begin{equation}
    \sigma = \frac{e^{2}E_{F}}{4\pi \hbar^{2}\left(i\omega - \tau_{DP}^{-1} \right)}
    \label{optiTI}
\end{equation}
where $E_{F}$ is the Fermi energy of the surface states, $\tau_{DP}$ is the relaxation time of the Dirac plasmon, and $e$ is the electron charge. We note that the hybridized states at the interface between a TI and another material (here the AFM) may have an impact on the carrier density at the interface \cite{To2022,Stauber2017} and make it different than that on the surface of a pristine TI layer. However, for simplicity we neglect this effect and assume the same optical conductivity expression for the surface of the TI and the interface between the TI and the AFM. In other words, in the following $\sigma_{0} \equiv \sigma_{1} \equiv \sigma$ as given by Eq. \ref{optiTI}. 

We note that in the long-wavelength limit $\left(k_{x}d_{TI} \ll 1 \right)$, the analytical expression for the surface plasmon-phonon-polariton in the TI thin film was derived in \cite{Stauber2017}

\begin{equation}
    \omega_{TI_{+}}^{2} = \frac{v_{F}\sqrt{2\pi n_{2D}}e^{2}}{\varepsilon_{0}h}\frac{k_{x}}{\varepsilon_{top} + \varepsilon_{bot}+k_{x}d_{TI}\varepsilon_{TI}}
    \label{TIdis}
\end{equation}
and 
\begin{equation}
    \omega_{TI_{-}}^{2} = \frac{2 \varepsilon_{0}\varepsilon_{TI}hv_{F}+e^{2}\sqrt{2\pi n_{D}}d_{TI}}{\sqrt{4\varepsilon_{0}^{2}\varepsilon_{TI}^{2}h^{2}v_{F}^{2}+2\varepsilon_{0}\varepsilon_{TI}e^{2}\sqrt{2 \pi n_{D}}d_{TI}}}k_{x}^{2}
\end{equation}
where the subscripts $TI_{+}$ and $TI_{-}$ stand for the optical and acoustic mode, respectively. Here $v_{F}$ is the Fermi velocity for the Dirac plasmon in the TI; $n_{2D}$ is the sheet carrier concentration of the entire TI thin film, including the contribution from both surfaces; $\varepsilon_{top}$, $\varepsilon_{bot}$ and $\varepsilon_{TI}$ are the permittivity of the top and bottom dielectric media and the TI, respectively; $k_{x}$ is the in-plane wave vector; and $d_{TI}$ is the thickness of the TI layer. In this work, we focus on the optical mode of the surface plasmon polariton in the TI because only this mode can be excited in a traditional optical experiment since the acoustic mode does not have any contribution in the optical dipole matrix element \cite{Ginley2018}. 

We also consider three AFM candidate materials, NiO, MnF$_2$ and FeF$_2$, which all support THz frequency magnons \cite{Rezende2019}. We note that the Néel temperature of NiO is about 523~K while that of FeF$_2$ and MnF$_2$ are, respectively, 78~K and 67~K \cite{Rezende2016}. We therefore imagine that the NiO sample could be studied experimentally at room temperature while the FeF$_2$ and MnF$_2$ samples would be investigated at low temperature. We assume that the samples are below their Néel temperatures in the calculations we conduct here. The frequency-dependent permeability of these AFMs is given given by Eq. \ref{mu}. The magnetic parameters and magnon frequencies of each AFM are listed in Table \ref{tableafm}. The permittivity of these AFMs are taken to be the same, with the characteristic value $\varepsilon^{AFM}=5.5$.

In order to gain physical insight into the formation of surface DPPMP modes in these TI/AFM structures, in parts \ref{DPMP1} and \ref{DPMP2} we consider ideal materials in which we set the scattering rate of all excitations to zero. In part \ref{DPMP3} we study how the non-zero realistic line widths of these excitations influence the coupling between a TI and AFM and the formation of hybridized states. 

\subsection{Formation of surface Dirac plasmon-phonon-magnon polaritons in TI/AFM structures}\label{DPMP1}
\begin{figure}[h]
\centering
    \includegraphics[width=.4\textwidth]{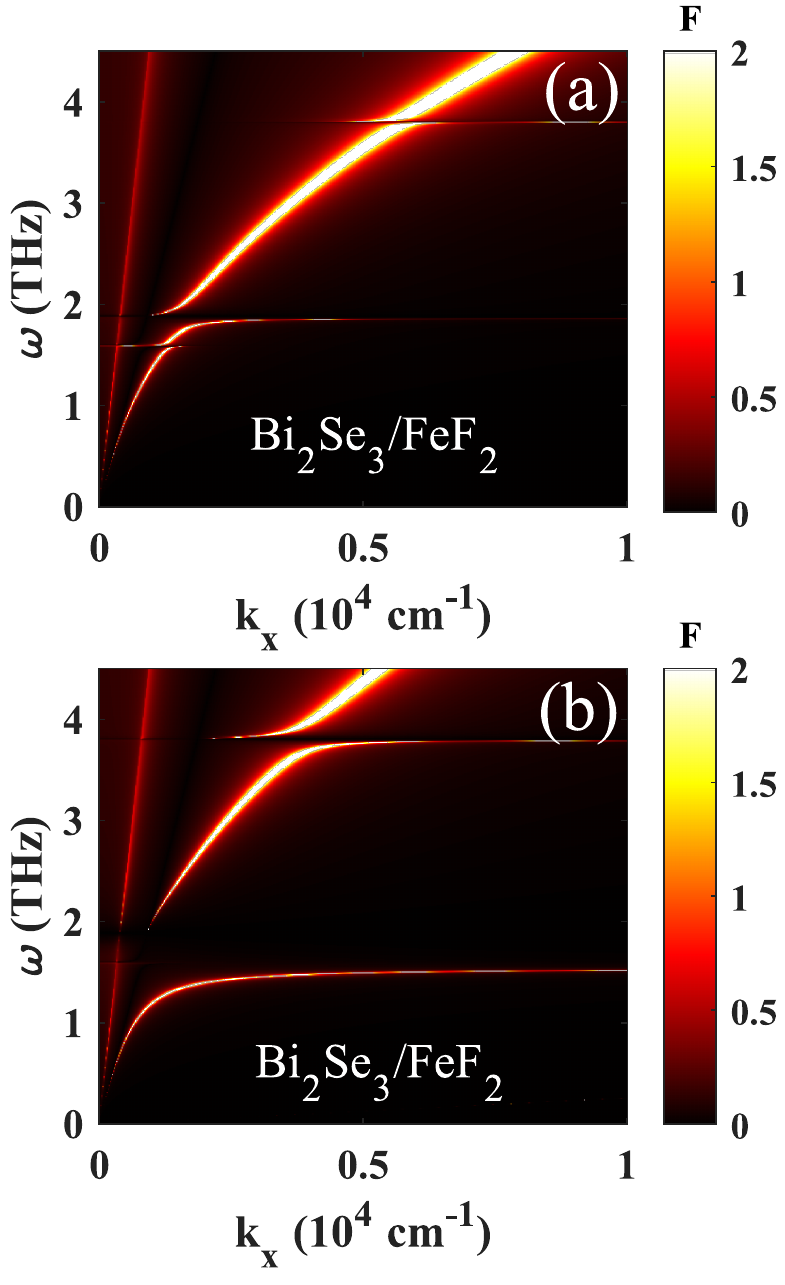}
 \caption{Surface DPPMP dispersion in Bi$_{2}$Se$_{3}$/FeF$_2$ structure calculated using Eq. \ref{TM_surface_p} with the Fermi energy of Dirac plasmon on its surface $E_{F}=1~eV$ and the thickness of TI layer (a) $d_{TI}=10~nm$ and (b) $d_{TI}=200~nm$.}
  \label{FIG4}
\end{figure}

\begin{figure}[htb]
\centering
    \includegraphics[width=.4\textwidth]{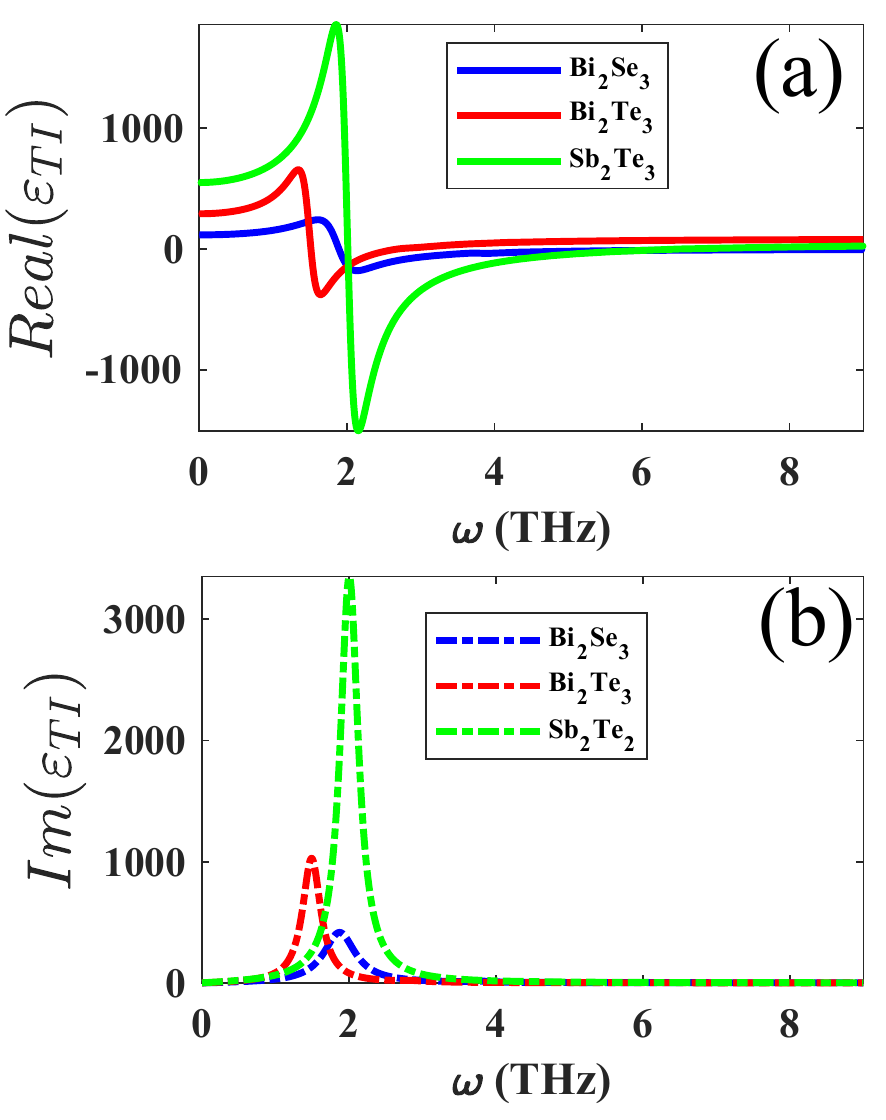}
 \caption{Frequency dependent dielectric function of topological materials for corresponding Bi$_2$Se$_3$ (blue), Bi$_2$Te$_3$ (red) and Sb$_2$Te$_3$ (green) with (a) the real part and (b) the imaginary part.}
  \label{FIG9}
\end{figure}

We begin with the dispersion relations of surface plasmon-phonons in a Bi$_2$Se$_3$ TI layer interacting with a magnon in a FeF$_2$ AFM. In Fig. \ref{FIG4} we plot the $F$ as given by Eq. \ref{TM_surface_p} as  a function of frequency $\omega$ and in-plane wave vector $k_{x}$. We apply this technique to Bi$_2$Se$_3$/FeF$_2$ bilayers with two different thicknesses of Bi$_2$Se$_3$ film: (a) $d_{TI}=10~nm$ and (b) $d_{TI}=200~nm$. In both cases we use a Fermi energy of the Dirac plasmon $E_{F}=1~eV$. The color in Fig. \ref{FIG4} represents the magnitude of the function $F$ whose maxima reveal the dispersion of the surface DPPMP. In Fig. \ref{FIG4}(a), for a rather thin Bi$_2$Se$_3$ layer, we observe the formation of a surface DPPMP through anti-crossings around 1.5~THz, 2~THz and 4~THz, i.e.~where the TI plasmon becomes degenerate with, respectively, the energies of the magnon in the FeF$_2$ material, the $\alpha$-phonon in the Bi$_2$Se$_3$, and the $\beta$-phonon in the Bi$_2$Se$_3$. 

As shown in Fig. \ref{FIG4}(b), when the thickness of the Bi$_2$Se$_3$ layer is increased to $d_{TI}=200~nm$ we continue to see the contributions from $\alpha$- and $\beta$-phonons in the dispersion, but the signature of the interaction between the Bi$_2$Se$_3$ and FeF$_2$ layer at 1.59~THz disappears. This is because the dispersion of the upper branch of the surface-plasmon-phonon polariton in a bare TI layer described by Eq. \ref{TIdis} blueshifts to above 2~THz with increasing Bi$_2$Se$_3$ thickness due to the negative real part of its permittivity in this domain. In contrast, the lower mode, below 2~THz, redshifts because the real part of the permittivity of Bi$_2$Se$_3$ is positive in this region. This redshift of the lower branch mode below 2~THz causes this mode to no longer intersect with the magnon resonant frequency at 1.59~THz. As a result, the anti-crossing associated with TI/AFM interaction cannot be observed in Fig. \ref{FIG4}(b). A similar effect is seen in all three types of TI materials considered in this work; Fig. \ref{FIG9} shows the frequency-dependent dielectric function of all three TIs, with the transition from positive to negative values of the real part of $\varepsilon_{TI}$ at around 2~THz. This analysis indicates an upper bound for the thickness of TI layer in which one could observe the coupling between a TI and an AFM. Specifically, in the case of Bi$_2$Se$_3$/FeF$_2$ structure, the thickness of the TI layer should not exceed 120~nm. 

\subsection{Impact of material parameters and Fermi energy on the coupling strength}
\label{DPMP2}
\begin{figure}[htb]
\centering
    \includegraphics[width=.4\textwidth]{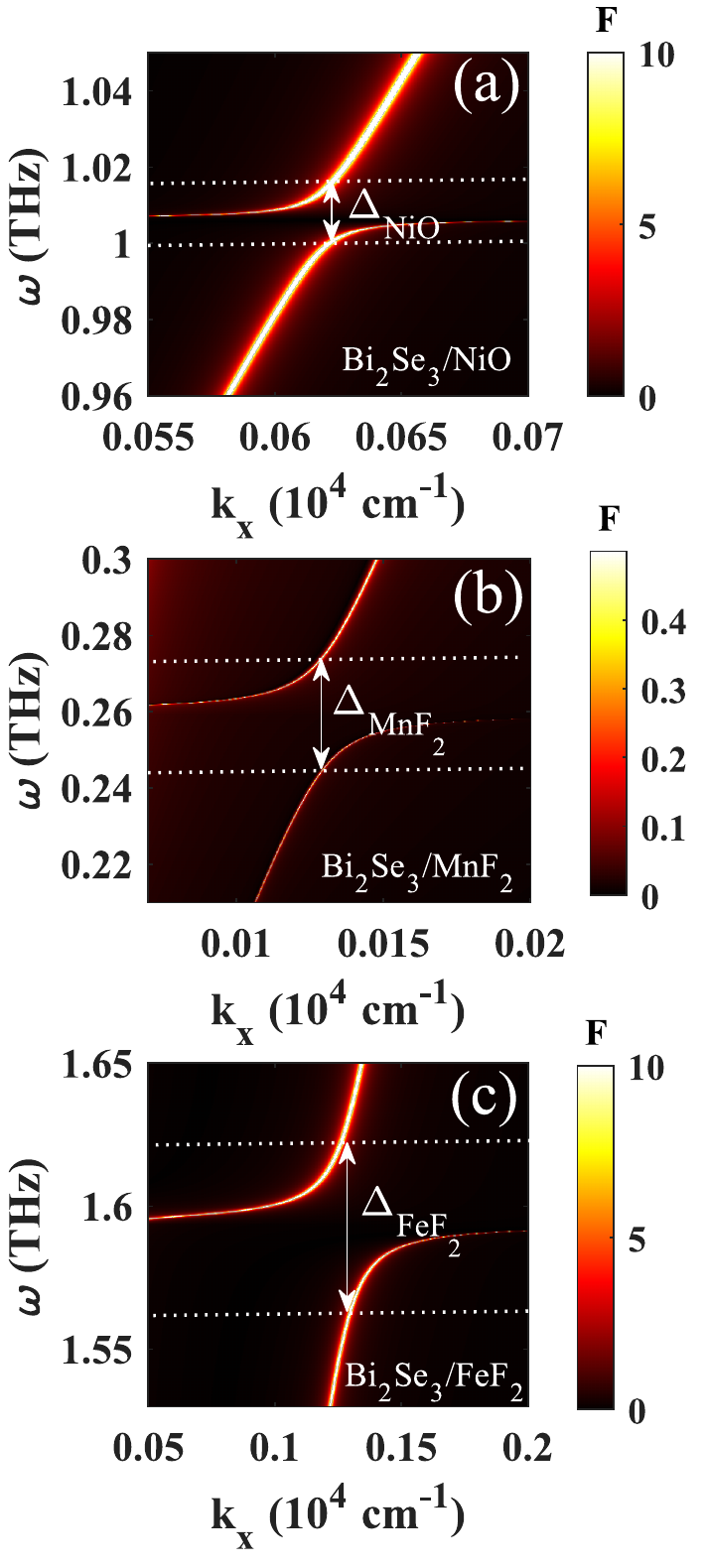}
 \caption{Surface DPPMP dispersion in the vicinity of magnon frequency calculated using Eq. \ref{TM_surface_p} for (a) Bi$_{2}$Se$_{3}$/NiO , (b) Bi$_{2}$Se$_{3}$/MnF$_2$, and (c) Bi$_{2}$Se$_{3}$/FeF$_2$  bilayer structures. In all cases $d_{TI}=10~nm$ and $E_{F}=1~eV$.}
  \label{FIG11}
\end{figure}
We now consider the impact of material parameters and Fermi energy on the strength of the interaction between a TI and an AFM. In Fig. \ref{FIG11} we show the dispersion of the DPPMP around the magnon frequency for three different TI/AFM bilayers: (a) Bi$_2$Se$_3$/NiO, (b) Bi$_2$Se$_3$/MnF$_2$, and (c) Bi$_2$Se$_3$/FeF$_2$. The thickness of Bi$_2$Se$_3$ and Fermi energy of the Dirac plasmon in all of these calculations is $d_{TI}=10~nm$ and $E_{F}=1~eV$, respectively. One observes that the coupling strength, defined by the magnitude of the separation $\Delta$ between the upper and lower modes near the magnon frequency and at a specific in-plane wavevector around the maximum anti-crossing point, increases from Bi$_2$Se$_3$/NiO (Fig. \ref{FIG11}a: $\Delta_{NiO}\approx 0.015~THz$) to Bi$_2$Se$_3$/MnF$_2$ (Fig. \ref{FIG11}b: $\Delta_{MnF_{2}}\approx 0.028~THz$) and has the largest value for Bi$_2$Se$_3$/FeF$_2$ (Fig. \ref{FIG11}c: $\Delta_{FeF_{2}}\approx 0.07~THz$). This is because the anisotropy constant $K=\gamma^{2}MH_{a}$ (where $\gamma$ is the gyromagnetic ratio, M is the sublattice magnetization and $H_a$ is the anisotropy field) that determines the magnitude of the magnetic dipole in the AFM is the largest for FeF$_2$ and smallest for NiO. We have also performed similar calculations for different TI materials combined with FeF$_2$ (data not shown) in which we used the same Fermi energy for the Dirac plasmon on the surfaces of all the TI films. These calculations tell us that the coupling strength between TIs and FeF$_2$ is almost independent of the change in TI materials. These data suggest that the interaction between a surface Dirac plasmon-phonon in a TI with a magnon in an AFM primarily depends on the amplitude of the magnetic dipole in the AFM material. In other words, one would expect, based on the data presented thus far, that a larger magnetic dipole for an AFM should lead to a stronger interaction between the surface DPP in a TI and the magnon in the AFM. However, as we will now show, this is not the case for the surface DPPMP in the TI/AFM bilayer at low Fermi energy, a regime typically at $E_{F}<0.4~eV$ for the materials investigated here.

\begin{figure}[h]
\centering
    \includegraphics[width=.4\textwidth]{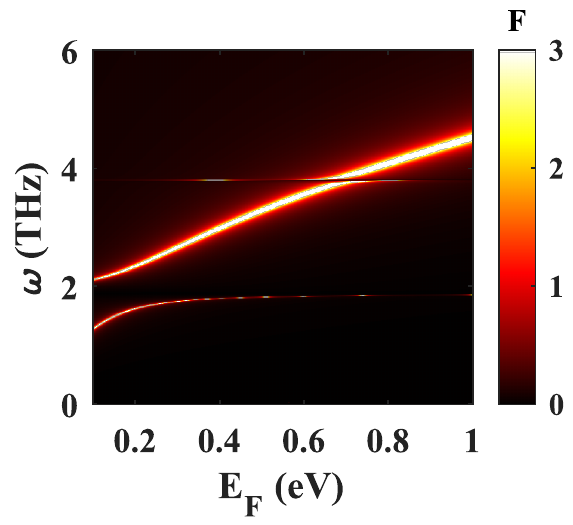}
 \caption{Surface DPPMP dispersion in Bi$_{2}$Se$_{3}$/FeF$_2$ bilayer structure calculated using Eq. \ref{TM_surface_p} as a function of varying Fermi energy of the Dirac surface plasmon. $d_{TI}=10~nm$ and the in-plane wave vector $k_{x}=0.8 \times 10^{4}~cm^{-1}$ are kept constant.}
  \label{FIG5}
\end{figure}

\begin{figure}[htb]
\centering
    \includegraphics[width=.4\textwidth]{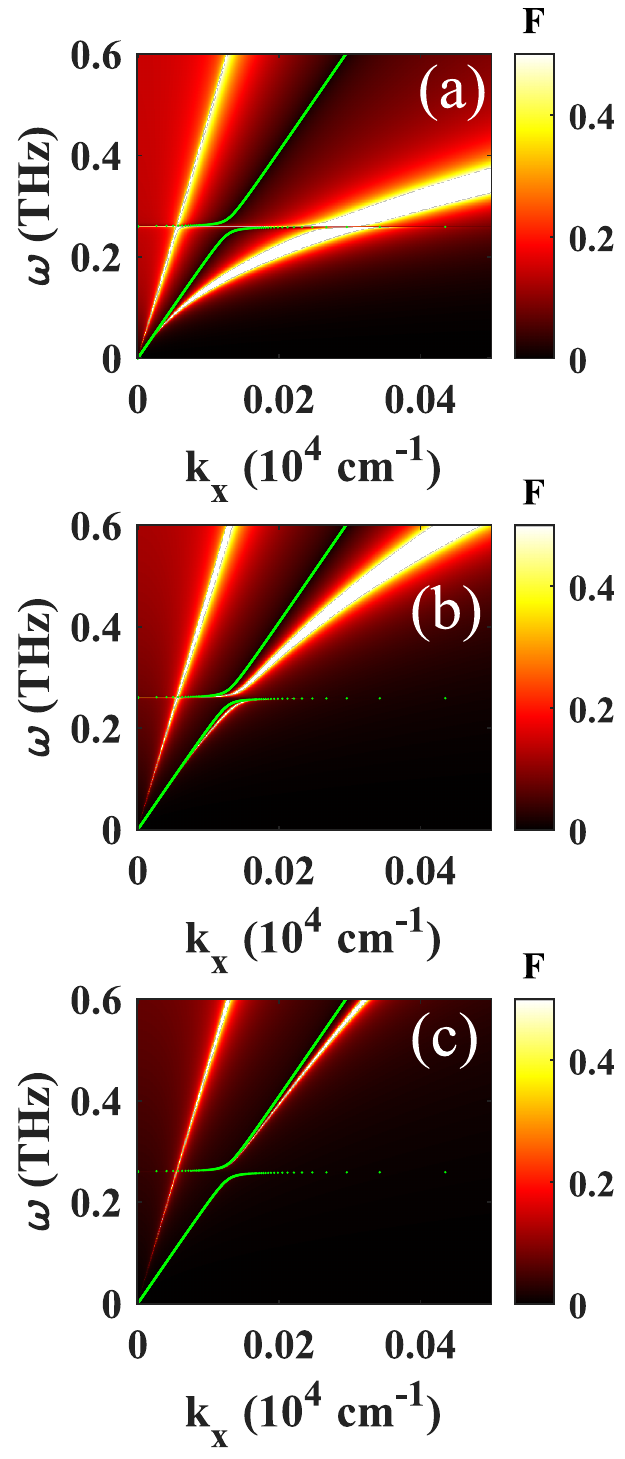}
 \caption{Surface DPPMP dispersion in Bi$_{2}$Se$_{3}$/MnF$_2$ structure calculated from Eq. \ref{TM_surface_p} with $d_{TI}=10~nm$, Fermi energy (a) $E_F=0.1~eV$, (b) $E_f=0.4~eV$, and (c) $E_f=1~eV$. The green dotted line in all figures is the dispersion of the bulk magnon polariton mode in the MnF$_2$. }
  \label{FIG13}
\end{figure}

\begin{figure}[h]
\centering
    \includegraphics[width=.4\textwidth]{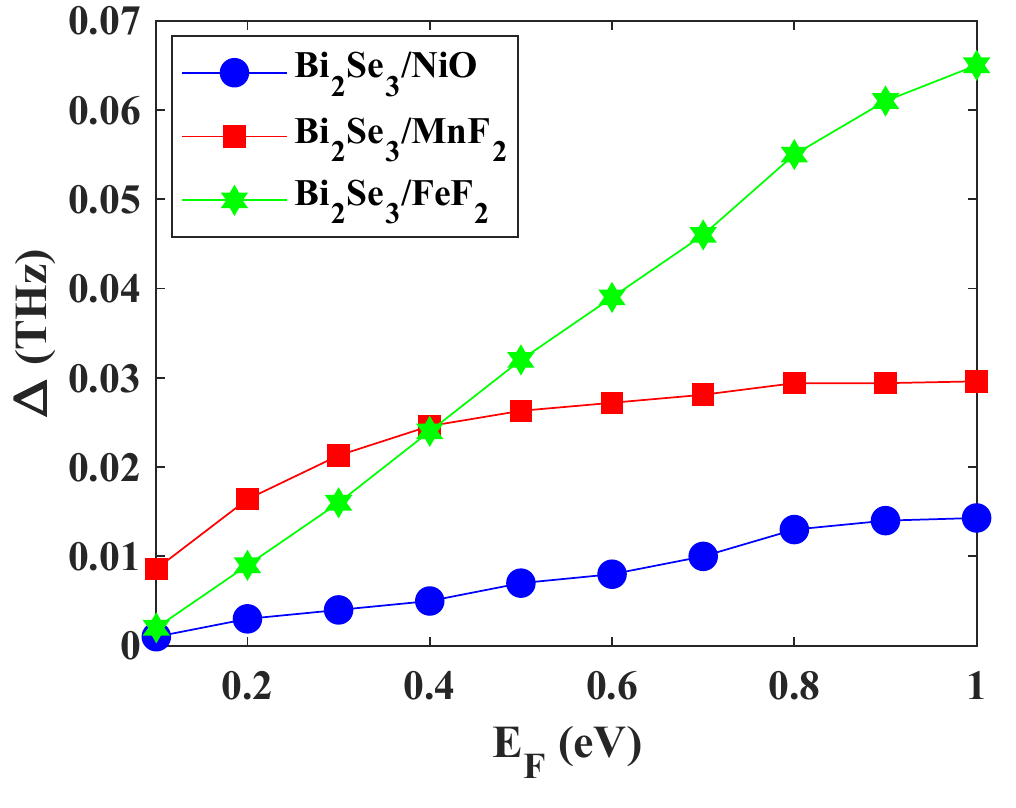}
 \caption{TI-AFM coupling strength in the Bi$_{2}$Se$_{3}$/FeF$_2$, Bi$_{2}$Se$_{3}$/NiO, and Bi$_{2}$Se$_{3}$/MnF$_2$ bilayer structures as a function of the Fermi energy of the Dirac plasmon in the TI film. In all cases $d_{TI}=10~nm$. Note that we use the technique presented in our previous work\cite{To2022} to extract the strength of the coupling between the TI and the AFM that we call $\Delta$; the magnitude of $\Delta$ is schematically depicted in Fig. \ref{FIG11}. For each bilayer the magnitude of $\Delta$ is extracted at the maximum anti-crossing (around 1.6~THz for Bi$_{2}$Se$_{3}$/FeF$_2$, at 1~THz for Bi$_{2}$Se$_{3}$/NiO, and around 0.26~THz for Bi$_{2}$Se$_{3}$/MnF$_2$).}
  \label{FIG6}
\end{figure}

We now investigate the influence of the Fermi level of surface states in the TI layer on the coupling between the TI and AFM. In Fig.~\ref{FIG5} we plot the energy of the surface DPPMP modes in a Bi$_{2}$Se$_{3}$/FeF$_2$ bilayer as a function of the Fermi energy of the TI surface states from 0 to 1~eV. We simply observe the blueshift of all polariton branches because the higher Fermi level leads to more electrons participating in the surface mode and that shifts the dispersion to higher frequency per Eq. \ref{TIdis}. A direct consequence of this blueshift is that increasing Fermi level from zero will always shift the Dirac plasmon-phonon-polariton closer toward resonance with the magnon polariton mode, thereby increasing the coupling strength. To understand how these shifts affect the coupling strength, in Fig.~\ref{FIG13} we plot the dispersion relations for the DPPMP in the Bi$_{2}$Se$_{3}$/MnF$_2$ bilayer for three different Fermi energies (a) $E_F=0.1~eV$, (b) $E_f=0.4~eV$, and (c) $E_f=1~eV$. We plot in green the dispersion of the bulk magnon polariton mode in the MnF$_2$, which is given by $k^2=\frac{\varepsilon^{AFM}\mu^{AFM}\omega^{2}}{c^2}$. For low Fermi level ($E_F=0.1~eV$, Fig.~\ref{FIG13}a) the Dirac plasmon-phonon-polariton in the TI crosses the MnF$_2$ magnon resonance frequency $\Omega_{0}=0.26~THz$ at approximately $k_{x}=0.028 \times 10^{4}~cm^{-1}$. The magnitude of the anti-crossing ($\Delta$, which is a measure of the strength of the coupling) is extremely small and barely visible in Fig.~\ref{FIG13}a. This occurs because the surface Dirac plasmon-phonon-polariton in the TI thin film at low Fermi energy is relatively far out of resonance with the magnon polariton mode in the AFM represented by the steepest green line in Fig.~\ref{FIG13}a. Being relatively far from resonance reduces the magnon contribution to the hybridized DPPMP state, reducing the coupling strength. In contrast, for slightly higher Fermi level ($E_f=0.4~eV$, Fig.~\ref{FIG13}b) the Dirac plasmon-phonon-polariton in the TI crosses the magnon resonance frequency at approximately $k_{x}=0.014 \times 10^{4}~cm^{-1}$. The magnitude of the anti-crossing ($\Delta$) is significantly larger because the Dirac plasmon-phonon-polariton in the TI are very close to the resonance point with magnon polariton in the AFM at $\omega=0.26~THz$ and $k_{x}=0.013 \times 10^{4}~cm^{-1}$ (i.e. the anticrossing feature in the green line). A further blueshift in the Dirac plasmon-phonon-polariton in the TI caused by a further increase in the Fermi Level ($E_f=1~eV$, Fig.~\ref{FIG13}c) does not further increase the magnitude of the anti-crossing ($\Delta$). To understand this saturation more clearly, in Fig.~\ref{FIG6} we plot $\Delta$ as a function of Bi$_{2}$Se$_{3}$ Fermi level for bilayers constructed with all three candidate AFM materials. The magnitude of $\Delta$ for the Bi$_{2}$Se$_{3}$/MnF$_2$ structures initially increases with increasing Fermi level and then saturates for $E_{F}>0.4~eV$ as described above with reference to Fig.~\ref{FIG13}. For similar reasons, the magnitude of $\Delta$ for the Bi$_{2}$Se$_{3}$/NiO structure initially increases and then saturates for $E_{F}>0.8~eV$. In contrast, $\Delta$ for the Bi$_{2}$Se$_{3}$/FeF$_2$ structure increases monotonically with increasing Fermi energy across the range of $E_{F}$ we consider here. The increasing threshold for saturation of $\Delta$ is directly related to the magnon frequency in the AFM. MnF$_2$ has the lowest magnon frequency and thus the lowest saturation $E_{F}$. FeF$_2$ has the highest magnon frequency and thus the highest saturation $E_{F}$ (above the range of $E_{F}$ considered here). Taken together, this analysis suggests that one way to tune the magnitude of the coupling between a TI and an AFM is by gating the electrons on the surface of the TI thin film.

\subsection{The role of linewidth}
\label{DPMP3}
\begin{figure}[h]
\centering
    \includegraphics[width=.37\textwidth]{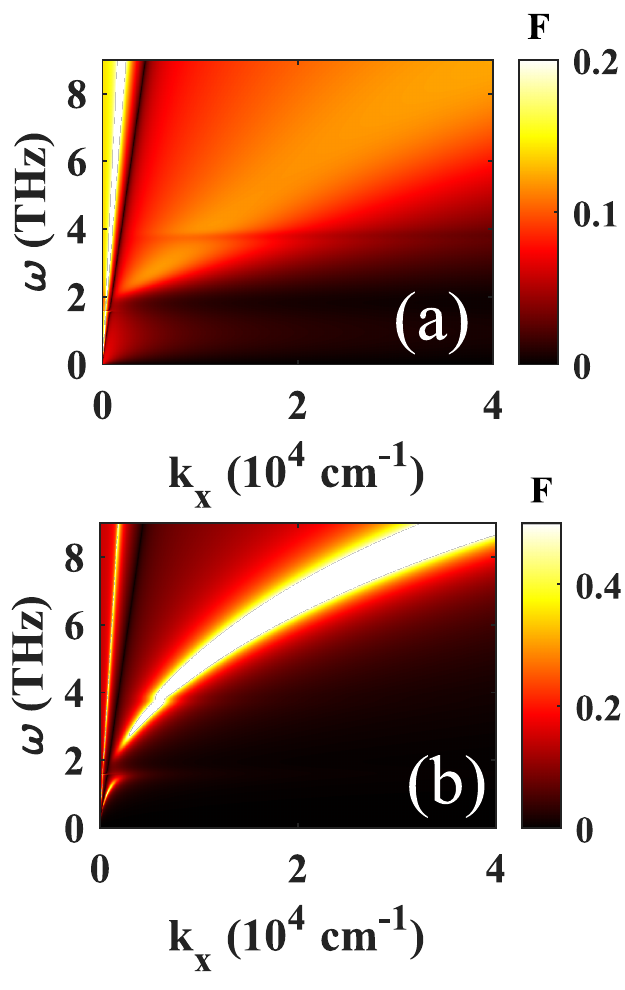}
 \caption{Surface DPPMP dispersion in Bi$_{2}$Se$_{3}$/FeF$_2$ structure calculated using Eq. \ref{TM_surface_p}. For both calculations the Fermi energy of the Dirac plasmon is $E_{F}=0.7~eV$, the thickness of the TI layer is $d_{TI}=80~nm$, and the linewidths of the $\alpha$- and $\beta$- phonon in the TI and the magnon in the AFM are set at the realistic values listed in Tables \ref{tableTI} and \ref{tableafm}. The linewidth of the Dirac plasmon is (a) $\Gamma_{D}=\frac{1}{\tau_{DP}}=10~THz$ and (b) $\Gamma_{D}=\frac{1}{\tau_{DP}}=0.1~THz$.}
  \label{FIG12}
\end{figure}
Thus far we have been considering ideal materials in which we neglected all loss rates by setting the linewidths to zero. This has allowed us to understand much of the physics underlying the emergence of hybridized states due to strong coupling. We will now consider realistic material parameters by including dissipative effects via non-zero linewidths. We start by considering the influence of scattering rate (linewidth) on the dispersion of the surface DPPMP for the Bi$_{2}$Se$_{3}$/FeF$_2$ structure. In Fig.~\ref{FIG12} we plot the energies of the surface DPMP modes for $d_{TI}=80~nm$, $E_{F}=0.7~eV$, and realistic values of the loss rates in the system described by the linewidths of the $\alpha$- and $\beta$- phonons in the TI and the magnon in the AFM as given in Tables \ref{tableTI} and \ref{tableafm}. The linewidth (relaxation time) of the Dirac plasmon for the calculation in Fig. \ref{FIG12}(a) is $\Gamma_{D}=\frac{1}{\tau_{DP}}=10~THz$ ($\tau=0.1~ps$). For Fig. \ref{FIG12}(b) the linewidth (relaxation time) is $\Gamma_{D}=\frac{1}{\tau_{DP}}=0.1~THz$ ($\tau=10~ps$). When the relaxation time of the Dirac plasmon is small [Fig.~\ref{FIG12}(a)], the polariton branch below 2~THz is invisible and there is no anti-crossing signature in the vicinity of $\omega\sim1.6~THz$. In contrast, for an increased relaxation time for the Dirac plasmon [Fig.~\ref{FIG12}(b)], the polariton branch below 2~THz becomes clear and the anti-crossing signature in the vicinity of $\omega\sim1.6~THz$ becomes apparent. This behavior of the polariton branch below 2~THz is due to the dominance of the TI surface states at the low frequency; at high frequency the $\alpha$- and $\beta$- phonons with smaller linewidths (see Table \ref{tableTI}) play an important role. The line width of the AFM is very small in comparison to that of TI and therefore has less impact on the whole dispersion relation for the DPPMP. These calculations imply that the coupling between TI and AFM can only be observed with extremely high TI quality (small linewidth). Specifically, for Bi$_2$Se$_3$ our calculation predicts that the linewidth of the Dirac plasmon on the surface of Bi$_2$Se$_3$ should not exceed 1~THz.

\begin{figure}[h]
\centering
    \includegraphics[width=.4\textwidth]{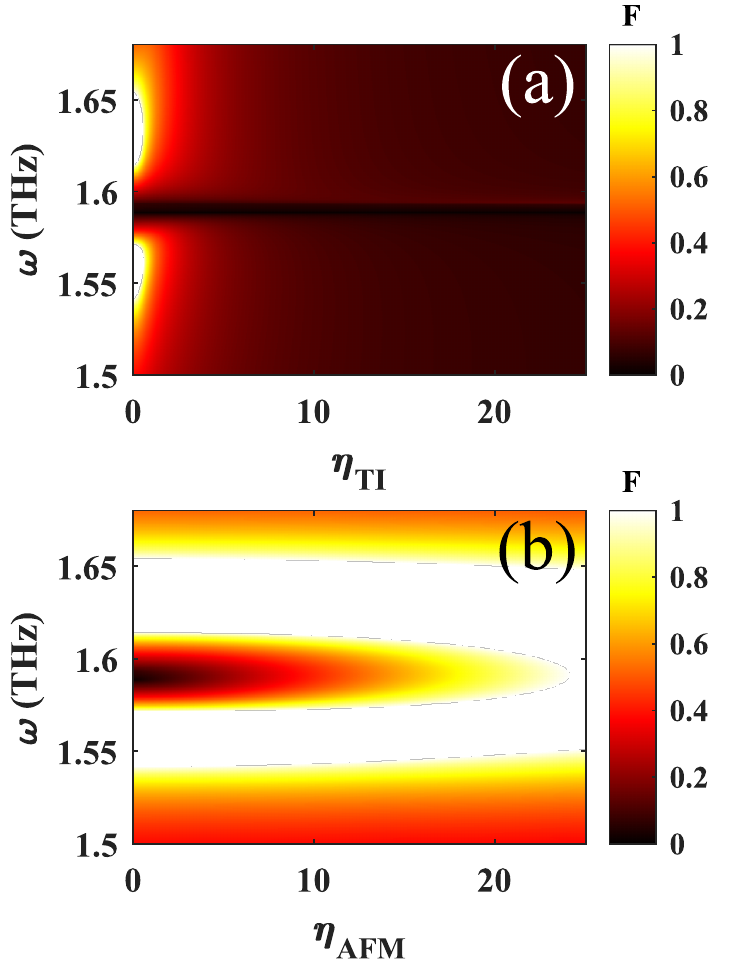}
 \caption{Surface Dirac plasmon-phonon-magnon polariton dispersion in Bi$_{2}$Se$_{3}$/FeF$_2$ structure versus the variations of TI's linewidth (a) and AFM linewidth (b) calculated from Eq. \ref{TM_surface_p} with the Fermi energy of Dirac plasmon on its surface $E_{F}=1~eV$ and the thickness of TI layer $d_{TI}=10~nm$ at a fixed in-plane wave vector $k_{x}=0.13\times10^{4}~cm^{-1}$.}
  \label{FIG7}
\end{figure}

Finally, we consider the effect of scattering rate on the coupling strength. For convenience and clarity, when we consider the effect of TI loss rate we will assume zero loss (zero linewidth) for the AFM and vice versa. Computationally, we change the scattering loss rate in the TI by adding a multiplicative factor $\eta_{TI}$ to the initial values of the linewidth of the $\alpha$- and $\beta$- phonons (listed in Table \ref{tableTI}) and the Dirac plasmon in the TI layer, which has initial value $\Gamma_{0}=0.1~THz$. In Fig. \ref{FIG7}(a) we plot the dispersion at a fixed in-plane wave vector of $k_{x}=0.1\times 10^{4}~cm^{-1}$ as a function of $\eta_{TI}$. We find that the scattering loss rate in the TI material has almost no impact on the strength of the coupling between the TI and the AFM as indicated by the persistently-observable splitting between the upper mode at around 1.64~THz and lower mode at 1.55~THz. We note, however, that the two branches of the polariton become progressively fainter upon increasing the linewidth of the TI, as discussed previously. In Fig. \ref{FIG7}(b) we present an analogous plot as a function of $\eta_{AFM}$. We find that the upper and lower polariton modes merge into a single polariton when $\eta_{AFM}\sim25$, i.e. for a factor or 25 increase in the linewidth of the magnon in the AFM relative to the base value listed in Table \ref{tableafm}. This merge into a single polariton occurs because when linewidth in the AFM exceeds the coupling strength (anti-crossing) and provides a benchmark for the AFM linewidth that would require to be achieved to create an experimentally-observable coupling between the TI and the AFM. Together with the above analysis, this result shows the critical importance of extremely high quality samples for any experimental study of the coupling between a TI and an AFM such as those  theoretically explored here.

\section{Conclusions}
\label{conc}
In summary, we have presented a robust method for investigating the surface polariton modes in a heterostructure by using the scattering and transfer matrix method with proper boundary conditions. We then apply this technique to systematically study the interaction between a TI and an AFM in a TI/AFM bilayer mediated by both the electric and magnetic degrees of freedom. In particular, we explore the limits for reaching the strong coupling regime evidenced by the formation of DPPMP hybridized states and emergence of an anti-crossing. Our calculations predict an upper bound for the thickness of the TI layer that could be used to experimentally observation such strong coupling. We quantify the dependence of the coupling strength on the magnetic dipole and linewidth of the magnon, which are intrinsic properties of the AFM material. We also quantify the dependence of the coupling on the Fermi energy, and hence the carrier concentration, of the Dirac plasmon on the surface of the TI and at the interface between the TI and the AFM. We find that the strength of the interaction between a TI and an AFM can be tuned via gating the electrons on the surface of the TI. We also find that the saturation of the coupling strength primarily depends on the magnetic dipole of the AFM material. Finally, we show that extremely high material quality is essential to experimentally observing strong coupling between the TI and the AFM. 

\begin{acknowledgments}
This research was primarily supported by NSF through the University of Delaware Materials Research Science and Engineering Center, DMR-2011824.
\end{acknowledgments}

\bibliography{Ref}% Produces the bibliography via BibTeX.

%\begin{thebibliography}{99}

%\end{thebibliography}

\end{document}